\documentclass[RNAAS]{aastex63}

\shortauthors{Zhao et al.}
\graphicspath{{./}{figures/}}

\begin{document}
\title{Estimating the black hole spin for the X-ray binary MAXI J1820+070}

\correspondingauthor{Xueshan~Zhao,Lijun~Gou}
\email{xszhao@nao.cas.cn,lgou@nao.cas.cn}

\author{Xueshan Zhao}
\affiliation{Key Laboratory for Computational Astrophysics, National Astronomical Observatories, Chinese Academy of Sciences,\\
Datun Road A20, Beijing 100012, China}
\affiliation{School of Astronomy and Space Sciences, University of Chinese Academy of Sciences, Datun Road A20, Beijing 100049, China}
 
 \author{Lijun Gou}
\affiliation{Key Laboratory for Computational Astrophysics, National Astronomical Observatories, Chinese Academy of Sciences,\\
Datun Road A20, Beijing 100012, China}
\affiliation{School of Astronomy and Space Sciences, University of Chinese Academy of Sciences, Datun Road A20, Beijing 100049, China}

\author{Yanting Dong}
\affiliation{Key Laboratory for Computational Astrophysics, National Astronomical Observatories, Chinese Academy of Sciences,\\
Datun Road A20, Beijing 100012, China}
\affiliation{Zhejiang Institute of Modern Physics, Department of Physics, Zhejiang University, 38 Zheda Road, Hangzhou 310027, China}

\author{Youli Tuo}
\affiliation{Key Laboratory of Particle Astrophysics, Institute of High Energy Physics, Chinese Academy of Sciences, 19B Yuquan Road, Beijing 100049,People’s Republic of China}

\author{Zhenxuan Liao}
\affiliation{Key Laboratory for Computational Astrophysics, National Astronomical Observatories, Chinese Academy of Sciences,\\
Datun Road A20, Beijing 100012, China}
\affiliation{School of Astronomy and Space Sciences, University of Chinese Academy of Sciences, Datun Road A20, Beijing 100049, China}

\author{Yufeng Li}
\affiliation{Key Laboratory for Computational Astrophysics, National Astronomical Observatories, Chinese Academy of Sciences,\\
Datun Road A20, Beijing 100012, China}
\affiliation{School of Astronomy and Space Sciences, University of Chinese Academy of Sciences, Datun Road A20, Beijing 100049, China}

\author{Nan Jia}
\affiliation{Key Laboratory for Computational Astrophysics, National Astronomical Observatories, Chinese Academy of Sciences,\\
Datun Road A20, Beijing 100012, China}
\affiliation{School of Astronomy and Space Sciences, University of Chinese Academy of Sciences, Datun Road A20, Beijing 100049, China}

\author{Ye Feng}
\affiliation{Key Laboratory for Computational Astrophysics, National Astronomical Observatories, Chinese Academy of Sciences,\\
Datun Road A20, Beijing 100012, China}
\affiliation{School of Astronomy and Space Sciences, University of Chinese Academy of Sciences, Datun Road A20, Beijing 100049, China}

\author{James F. Steiner}
\affiliation{Harvard-Smithsonian Center for Astrophysics, Cambridge, MA 02138, United States}

\begin{abstract}
MAXI J1820+070 is a newly-discovered black hole X-ray binary, whose dynamical parameters, namely the black hole mass, the inclination angle and the source distance, have been estimated recently. \emph{Insight}-HXMT have observed its entire outburst from March 14th, 2018. In this work, we attempted to estimate the spin parameter~$a_*$, using the continuum-fitting method and applying  a fully-relativistic thin disk model to the soft-state spectra obtained by \emph{Insight}-HXMT. It is well know that $a_*$ is strongly dependent on three dynamical parameters in this method, and we have examined two sets of parameters. Adopting our preferred parameters: $M$ = $8.48^{+0.79}_{-0.72}~M_\odot$, $i=63^\circ\pm3^\circ$ and $D=2.96\pm0.33$ kpc, we found a slowly-spinning black hole of $a_*=0.14 \pm 0.09$ ($1\sigma$), which give a prograde spin parameter as majority of other systems show. While it is also possible for the black hole to have a retrograde spin (less than 0) if different dynamical parameters are taken.
\end{abstract}

\keywords{\emph{Insight}-HXMT, black hole physics, X-rays:binaries, X-rays: individual (MAXI J1820+070)}

\section{introduction}
\label{into}
The spin is a crucial parameter of a black hole, which is helpful to understand the driving mechanism of the relativistic jets (\citealt{Blandford1977}), to explore the fundamental physic around the black hole (\citealt{Wong2012}), to test the predictions of general relativity (\citealt{Bambi2011}, \citealt{Bambi2013}, \citealt{Tripathi2020}). We usually use a dimensionless parameter $a_*$ to represent the spin, defining $a_{*} \equiv a/M = cJ/GM^2$, where $M$ and $J$ represent the black hole mass and angular momentum. In the traditional electromagnetic (EM) domain, we can only estimate the spin with some indirect techniques since the spin only makes itself notable via the general relativistic (GR) effect within a small region around the event horizon\footnote{It is noted that both the gravitational wave (\citealt{Abbott2020}) and Event Horizon Telescope (EHT) observations can directly constrain the black hole spin. The gravitational wave observations have provided spin constraints for tens of merging black hole systems. As to EHT, it can only resolve the event horizons for supermassive black holes rather than stellar mass black holes (\citealt{eht2019}). }. So far, there have developed two widely-used methods to measure the spin parameter of accreting stellar-mass black holes: (1) the continuum-fitting method, which models the shape of the thermal emission from the accretion disk (\citealt{Zhang1997}); (2) the reflection-fitting method, which models the red wing of the relativistically-broadened and asymmetric Fe K$\alpha$ line (\citealt{Fabian1989}, \citealt{Reynolds2003}).

The key process of the spin measurements is to estimate the radius of the inner disk $r_{\rm in}$ ($r_{\rm in} \equiv cR_{\rm in}/GM$). $R_{\rm in}$ is assumed to be equal to the radius of the innermost stable circular orbit $R_{\rm ISCO}$ (\citealt{Shafee2008a}, \citealt{Reynolds2008}, \citealt{Penna2010}, \citealt{Kulkarni2011}, \citealt{Noble2009}, \citealt{Noble2010}, \citealt{Noble2011}), which is a monotonic function of the dimensionless spin parameter $a_{*}$, decreasing from 6$R_{\rm g}$ ($R_{\rm g}$ is the gravitational radius, which is defined as $R_{\rm g} = GM/c^{2}$) to 1$R_{\rm g}$ as the spin increases from $a_{*}$ = 0 to $a_{*}$ = 1 (\citealt{Bardeen1972}).

MAXI J1820+070 is a newly-discovered transient source. Its optical counterpart, ASASSN-18ey, was discovered by the All-Sky Automated Survey for SuperNovae (ASAS-SN, \citealt{Shappee2014}) on March 6th, 2018 at R.A. = 18$^{\rm h}$20$^{\rm m}$21.$^{\rm s}$9 dec. = +07$^{\circ}$11$^{\prime}$07.$^{\prime \prime}$3 (J2000) (\citealt{Tucker2018}). In X-ray, it was found by the Monitor of All-sky X-ray Image (MAXI, \citealt{Matsuoka2009}) on March 11st, 2018(\citealt{Kawamuro2018}). Since discovery, the X-ray outburst of MAXI J1820+070 displayed a fast increase, and then a slow decay (MJD 58200-MJD 58290) in flux. The source underwent its first re-brightening (MJD 58290-MJD 58305) and then dropped sharply as transiting to the soft state. The source stayed in the soft state for over 2 months (till around MJD 58380). After around MJD 58400, the source faded away into quiescence. During the entire eruption, this source showed all of the standard accretion state (\citealt{Remillard2006}): the thermal dominant state (TD), or the high/soft state (HSS); the low/hard state (LHS); the steep power law (SPL) state (the so-called intermediate state in some literature). In order to ensure the reasonable application of the continuum-fitting method, we focus on spectra dominated by thermal accretion disk component avoiding the interference introduced by the strong Comptonization component. Based on the previous works, the continuum-fitting method is only reliably applied to a thin accretion disk, i.e., the bolometric Eddington-scaled luminosity $l = L(a_*,\dot M)/L_{\rm Edd} < 0.3$ (equivalent to the aspect ratio $H/R < 0.05$; \citealt{Shafee2008b}).

In the continuum-fitting method, one determines $r_{\rm in}$ by fitting the X-ray thermal continuum from the accretion disk to the Novikov-Thorne thin disk model (\citealt{Novikov1973}). As a nonrelativistic approximation, given that the disk luminosity $L \approx 2 \pi D^2F($cos$i)^{-1} \approx 4 \pi R_{\rm ISCO}^2T_{\rm eff}^4$ (where $F$, $T_{\rm eff}$, $D$, $i$ indicate the X-ray flux, the effective temperature, the distance to the source and the inclination angle), we have $r_{\rm ISCO}^2 \approx F/(2~T_{\rm eff}^4~($cos$i)^{-1}~D^2~M^{-2})$. Therefore three dynamical parameters, namely the black hole mass $M$, the disk inclination $i$ and the source distance $D$, are crucial for estimating the spin. As we know, $a_{*}$ is inversely proportional to $R_{\rm ISCO}$/$M$ (\citealt{Bardeen1972}), so that a higher mass will lead to a higher $a_{*}$. Besides, we can see that the higher the inclination or the distance is, the higher the $R_{\rm ISCO}/M$ is, hence the smaller the spin is. 

There have been systematic measurements to determine these three parameters. \cite{Torres2019} reported a mass function $f(M)\equiv$($M_1$sin$i)^3/(M_1+M_2)^2=5.18\pm0.15~M_{\odot}$ (where $M_1$ and $M_2$ indicate the masses of the black hole and the donor star), dynamically confirming a black hole harboring in this binary. Assuming a provisional mass ratio $q\equiv M_2/M_1=0.12$, they constrained the binary inclination to be $69^\circ \lessapprox i \lessapprox 77^\circ$ and derived a black hole mass in the range of 7-8~$M_\odot$. \cite{Atri2020} calculated the distance to be $D=2.96\pm0.33$ kpc via the measurement of the radio parallax \footnote{The radio parallax measurement of 0.348 $\pm$ 0.033 mas is consistent with the value of 0.31 $\pm$ 0.11 mas estimated by Gaia-DR2. The distance estimated from the Gaia parallax is 3.46$^{+2.18}_{-1.03}$ kpc based on an exponentially decreasing space density prior (\citealt{Gandhi2019})}, using the Very Long Baseline Array (\emph{VLBA}) and the European Very Long Baseline Interferometry (\emph{VLBI}) Network. Further they used the distance and estimated the jet inclination angle and the black hole mass to be  $63^\circ\pm3^\circ$ and $9.2\pm1.3~M_\odot$ (adopting $q=0.12$), respectively. A few months later, \cite{Torres2020} analyzed intermediate resolution optical spectroscopy, leading to a direct and accurate determination of $q=0.072\pm0.012$. Their constraint to the binary inclination is $66\fdg2< i < 80\fdg8$ based on the detection of eclipse and measurements of the accretion disk radius at the time of the optical spectroscopy, ignoring the disk vertical structure. The inclination angle derived in this way should represent the orbital inclination angle, and the corresponding black hole mass is $5.96~M_\odot < M < 8.06~M_\odot$.  Adopting $63^\circ\pm3^\circ$ (\citealt{Atri2020}), which could be consistent with the inclination angle of the inner accretion disk, the black mass would be $M$ = $8.48^{+0.79}_{-0.72}~M_\odot$, with uncertainties quoted at 1$\sigma$. For the continuum-fitting method, although it is typically assumed that the orbital inclination angle is aligned with the one for the inner accretion disk, it is found that there exists some misalignments in some systems (\citealt{Fragos2010}, \citealt{Walton2016}). Therefore, it is better to use the inclination angle which is more close to the one of the inner disk. In our case, we adopt $M$ = $8.48^{+0.79}_{-0.72}~M_\odot$, $i=63^\circ\pm3^\circ$ and $D=2.96\pm0.33$ kpc as our favored and primary parameter set to constrain the spin of MAXI J1820+070 via the continuum-fitting method. As a complement, we also discussed the spin parameter for the alternative parameter set in the discussion section. 

Many previous work have studied the characteristics of MAXI J1820+070, especially, the behaviour of inner disk radius (which implies the spin) of the black hole. \cite{Kara2019} performed spectra and temporal analysis on MAXI J1820+070, suggesting that as the hard state began to transit softward, the corona reduced in the spatial extent, while the inner disk radius was stable and approximately $\sim 2~R_{\rm g}$.  \cite{Bharali2019} fitted spectra obtained by Swift and NuSTAR in the hard state, constrained the inner disk radius and the disc inclination angle to be $5.1^{+1.0}_{-0.7} R_{\rm g}$ and $29.8^{+3.0}_{-2.7}$ $^\circ$ (3$\sigma$). By investigating the spectra evolution of the entire outburst using data from MAXI and Swift, \cite{Shidatsu2019} showed that the state transition occurred at its first re-brightening phase, suggesting that it is not determined by the mass accretion rate alone. They gave the constraint of the inner disk to be $R_{\rm in}$ = 77.9$\pm$1.0(D/3 kpc)(cos$i$/cos60$^{\circ}$)$^{-1/2}$ km during the high/soft state, with applying a combined correction factor of 1.18, for both the stress-free boundary condition and the color-temperature correction (\citealt{Kubota1998}). \cite{Buisson2019} analyzed data obtained by NuSTAR during the hard state, reporting a small stable inner radius, which implies a low-to-moderate-spin black hole. \cite{Fabian2020} discovered an excess emission between 6-10 keV during the soft state observed by NuSTAR and NICER, which can be well-modelled by an additional blackbody component {\rm BBODY}. They explained this excess to be the emission from the edge of the plunge region where matter begins to fall into the black hole. Both \cite{Fabian2020} and \cite{Buisson2019} reported that if the inner disk inclination lies between 30$^\circ$-40$^\circ$, then the spin lies in the range of $\pm$0.5. However, there has been no specific work to measure the spin of the black hole in MAXI J1820+070. Therefore, in this paper, we attempt to estimate the spin of this source by the continuum-fitting method.

This paper is organized as follows. In Section~\ref{observations}, we describe our observations and data reduction. In Section~\ref{analysis}, we report our spectral analysis and fit results. Discussion and conclusion are showed in Section~\ref{discussion}.

\section{observations and data reduction}
\label{observations}
The \emph{Hard X-ray Modulation Telescope} (HXMT, also named {\itshape Insight})\footnote{http://www.hxmt.org} was successfully launched on June 15th, 2017. There are three main detectors onboard \emph{Insight}-HXMT\footnote{http://www.hxmt.org/index.php/enhome/enabouthxmt/160-hard-x-ray-modulation-telescope}: the low energy X-ray telescope (LE, 1-15 keV, SCD, 384 cm$^2$); the medium energy X-ray telescope (ME, 5-30 keV, Si-Pin, 952 cm$^2$); the high energy X-ray telescope (HE, 20-250 keV, NaI(Tl)/CsI(Na), 5100 cm$^2$) (\citealt{Liu2019},~\citealt{Zhang2019},~\citealt{Chen2019},~\citealt{Cao2019}). As a collimated telescope, it has a negligible pile-up effect, which is suitable to observe bright source such as MAXI J1820+070. It began fixed-point observations for MAXI J1820+070 on March 14th, 2018 (MJD 58191), and monitored the whole outburst of this source until October 21st, 2018 (MJD 58412).

Spectra were extracted using {\itshape Insight}-HXMT Data Analysis Software (HXMTDAS) v2.02 following a standard procedure. Only spectra from events that belong to the small FoV (1.6$^\circ$ $\times$ 6$^\circ$ for LE and 1$^\circ$ $\times$ 4$^\circ$ for ME) were extracted. The background was measured by the blind detectors. Below 7 keV, the diffuse X-ray background is dominant while the particle background dominates above 7 keV (\citealt{Guo2020}, \citealt{Liao2020}). We estimated the level of the source contamination using HXMT Bright Source Warning Tool\footnote{See the proposal software page in http://proposal.ihep.ac.cn/soft/soft2.jspx, and see the detailed instructions in http://proposal.ihep.ac.cn/soft/soft2help.jspx}, finding no other interference source (with the flux ratio of the interference source to MAXI J1820+070 no more than 0.1) in the FoVs during observations. We simply used observations from LE and ME detectors, given that these instruments have already provided adequate energy coverage and that the extremely low net counts rate in HE. The spectra were rebinned using {\sc grppha} with at least 100 counts per new bin. We also added systematic errors due to uncertainties of the instrumental responses and background estimation: 1$\%$ for LE and 2$\%$ for ME (\citealt{Chen2019},~\citealt{Cao2019}). Spectral analysis was performed on {\sc Xspec} v12.10.1. In this work, we analyzed 2-10 keV for LE and 10-25 keV for ME.

\section{spectral analysis}
\label{analysis}

\subsection{Non-Relativistic Models}
\begin{figure}
\epsscale{1.0}
\centering
 \plotone{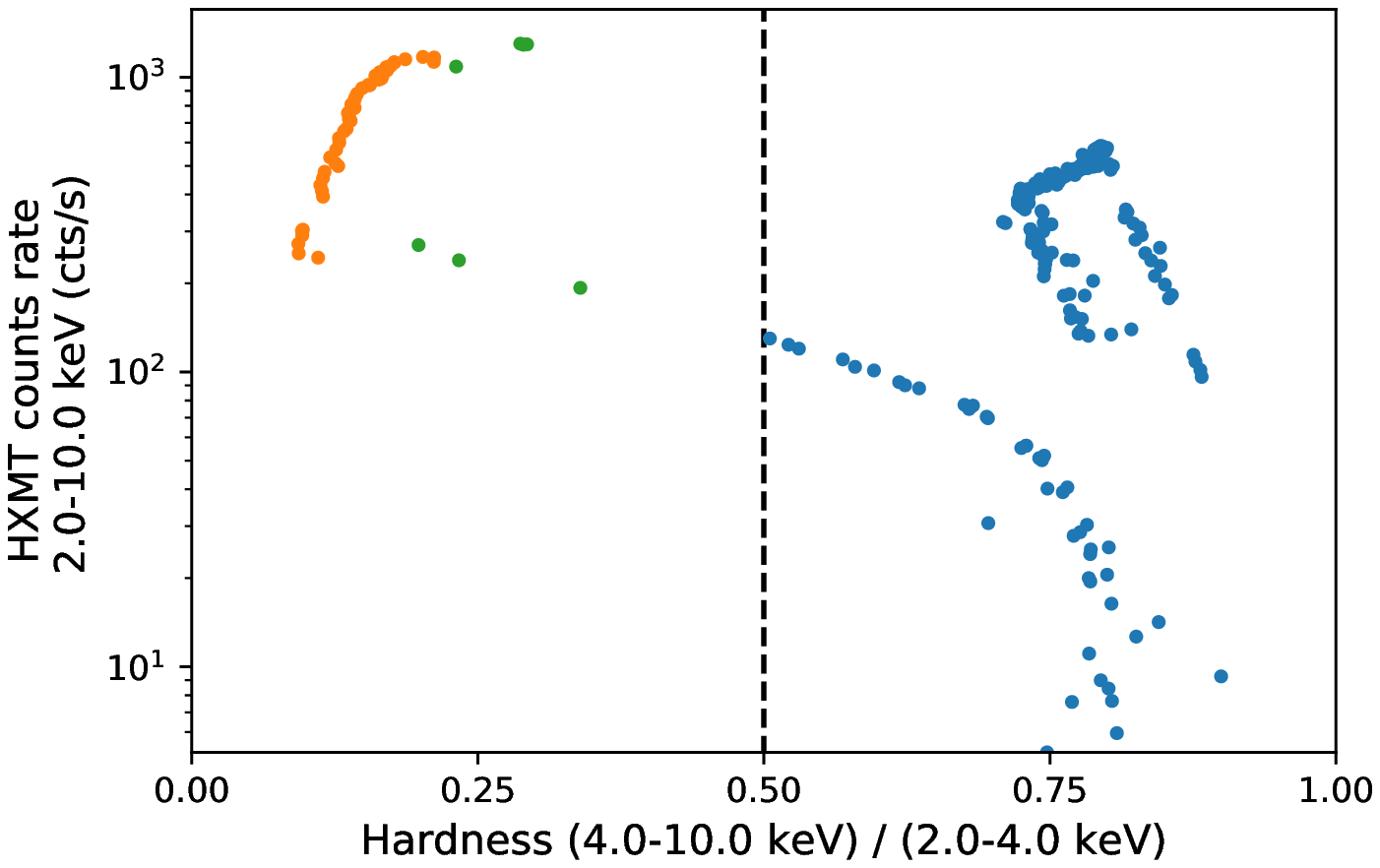}
 \plotone{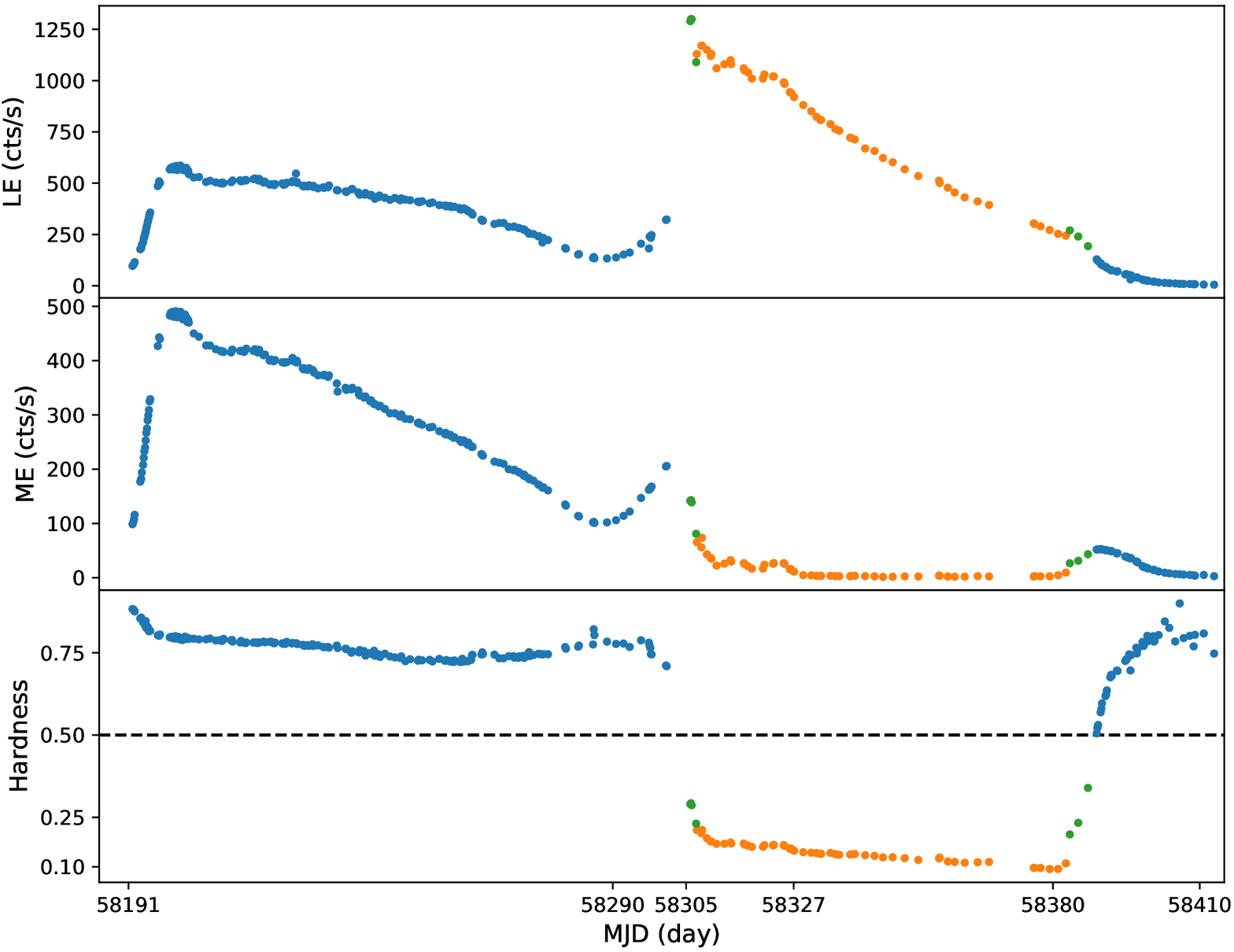}
\caption{\emph{Insight}-HXMT observations of MAXI J1820+070 covering its entire outburst. From top to bottom: the hardness-intensity diagram (HID), LE lightcurve, ME lightcurve, the evolution of the hardness ratio. The black dashed lines indicate the hardness ratio is 0.5. The blue points represent that the source is in the low/hard state, which are not analyzed in this paper. The orange and green symbols indicate observations with smaller and larger $f_{\rm sc}$, respectively. \label{fig:HXMT}}
\end{figure}

Figure~\ref{fig:HXMT} displays HXMT observations of MAXI J1820+070 during the 2018 outburst. As an empirical choice, we ignored spectra with hardness larger than 0.5, since our targets are the spectra dominated by thermal emission. 

We first applied a preliminary non-relativistic model on these spectra, which is expressed as {\sc constant*tbabs*simpl*diskbb} (hereafter, NR). The equivalent hydrogen column density in {\sc tbabs} was fixed to 0.15 $\times 10^{22}$~cm$^{-2}$ to account for the Galactic absorption (\citealt{Uttley2018}). Model {\sc constant} is used for coordinating calibration differences between the two detectors. We fixed the normalization of LE to 1, and the normalization of ME to 0.95 (Details will be addressed in Section~\ref{discussion}). In the empirical Comptonization model {\scshape simpl}, the parameter $f_{\rm sc}$ calculates the fraction of the thermal seed photons that are scattered into the power-law tail (\citealt{Steiner2009a}). \cite{Steiner2009b} suggested that the inner disk radius (i.e. the spin) remains stable to within a few percent as long as $f_{\rm sc} \lesssim 25\%$. Thereafter it became a widely-utilized spectral selection criteria in the continuum-fitting measurements of the spin. Because {\sc simpl} redistributes seed photons to both lower and higher energies where the response matrices of HXMT are limited, we extended the sampled energies to 0.1-100 keV on {\sc Xspec} (see the appendix of \citealt{Steiner2009a}). In the non-relativistic multicolors black body model {\sc diskbb}, two crucial parameters are estimated, namely the temperature ($T_{\rm in}$) and the radius ($R_{\rm in}$) of the inner accretion disk.

The fit results to NR model are listed in Table~\ref{diskbb}. For NR, It can be seen obviously that the $f_{\rm sc}$ of these spectra all satisfy $f_{\rm sc} \lesssim 25\%$, with the central values ranging from 1.4\% to 18.6\%. The temperature at inner disk radius $T_{\rm in}$ shows a small decline, lying in a range of  0.758-0.478 keV with the average value of 0.681 keV.  The normalization of {\sc diskbb} is defined as $(R_{\rm app}/D_{\rm 10})^2$cos$i$, where $R_{\rm app}$ is the apparent inner disk radius in unit of km. As defined in \cite{Kubota1998}, the realistic inner disk radius $R_{\rm in} = \xi f R_{\rm app}$, where $\xi$ is a correction factor for the stress-free inner boundary condition and $f$ is the spectral hardening factor. We assumed a combined correction factor of 1.18 (see \citealt{Kubota1998} for detail) following \cite{Shidatsu2019} in order to better compare with their results. Adopting $M$ = $8.48^{+0.79}_{-0.72}~M_\odot$, $i=63^\circ\pm3^\circ$ and $D=2.96\pm0.33$ kpc, we calculated the inner edge of the disk $R_{\rm in}$ in the unit of $R_{\rm g}$ (the gravitational radius $R_{\rm g} \equiv GM/c^2=12.55$~km for $M=8.48~M_\odot$). The best-fitting value of $R_{\rm in}$ varies slightly from 5.16$R_{\rm g}$ to 6.36$R_{\rm g}$. The averaged $R_{\rm in}$ is 5.58$R_{\rm g}$. The value of $R_{\rm in}$ estimated from {\sc diskbb} implies a small spin parameter. However, probably owing to the weak Comptonization component and the lower counts rate in E $>$ 10 keV, we were unable to constrain the photon index $\Gamma$ of SP25-52 (this will be addressed in later).


\subsection{Relativistic Models}

Next, we substituted {\sc diskbb} (\citealt{Mitsuda1984}, \citealt{Makishima1986}) with {\sc kerrbb2} (\citealt{McClintock2006}). The model is expressed as {\sc constant*tbabs*simpl*kerrbb2} (hereafter, RM). {\sc kerrbb2} is a fully-relativistic thin disk model and is the combination of {\sc bhspec} (\citealt{Davis2005}) and {\sc kerrbb} (\citealt{Li2005}). This model reads in a pair of look-up tables for the spectral hardening factor $f$ ($f \equiv T_{\rm col}/T_{\rm eff}$) estimated by {\sc bhspec} corresponding to two typical values of the viscosity parameter $\alpha$: 0.1 and 0.01. $a_*$ decreases slightly as $\alpha$ increases, so that throughout this work, we adopt $\alpha=0.01$ in order to make a more conservative limitation. Fit results for $\alpha=0.1$ are listed in Table~\ref{alpha}. A comparison of Tables 3 and 4 reveals that $\alpha=0.01$ gives systematically larger values of $a_*$, with a difference typically $\sim 0.05$. Specifically, referring to \cite{Shidatsu2019}, the parameter $\Gamma$ of SP25-52 is fixed to 2.5, a representative value for the soft state. For other spectra, we allowed $\Gamma$ to vary. 

The best-fit results are showed in Table~\ref{kerrbb2}. In RM, the values of $f_{\rm sc}$ in SP55-57 and SP61 (25.37\% $\pm$ 0.24\%, 25.21\% $\pm$ 0.24\%, 25.89\% $\pm$ 0.29\%, 27.78\% $\pm$ 0.21\%) are slightly above 25\%. And the value of $f_{\rm sc}$ in SP58-60 (14.79\% $\pm$ 0.22\%, 15.73\% $\pm$ 0.18\%, 18.75\% $\pm$ 0.20\%) are significantly larger than that of other 54 spectra. It means that the strength of the Comptonization component in SP1-54 is faint, contributing less than about 10\% of the total emission. The spin measurement using the continuum-fitting method based on those 54 spectra will be more robust. In fact, in SP55-61, MAXI J1820+070 underwent the state transition. \cite{Shidatsu2019} divided these 7 spectra into the intermediate state (IM). 

Therefore, we treated these spectra as our ``gold" spectra, mainly basing our analysis and error analysis on them although almost all the spectra meet the selection criteria $f_{\rm sc} \lesssim 25\%$. Meanwhile, SP55-61 (hereafter,``silver" spectra) were also analyzed just as reference. In Table~\ref{diskbb} and  Table~\ref{kerrbb2}, we used horizontal lines to distinguish the ``gold" spectra.

\begin{figure}
\epsscale{1.0}
 \plotone{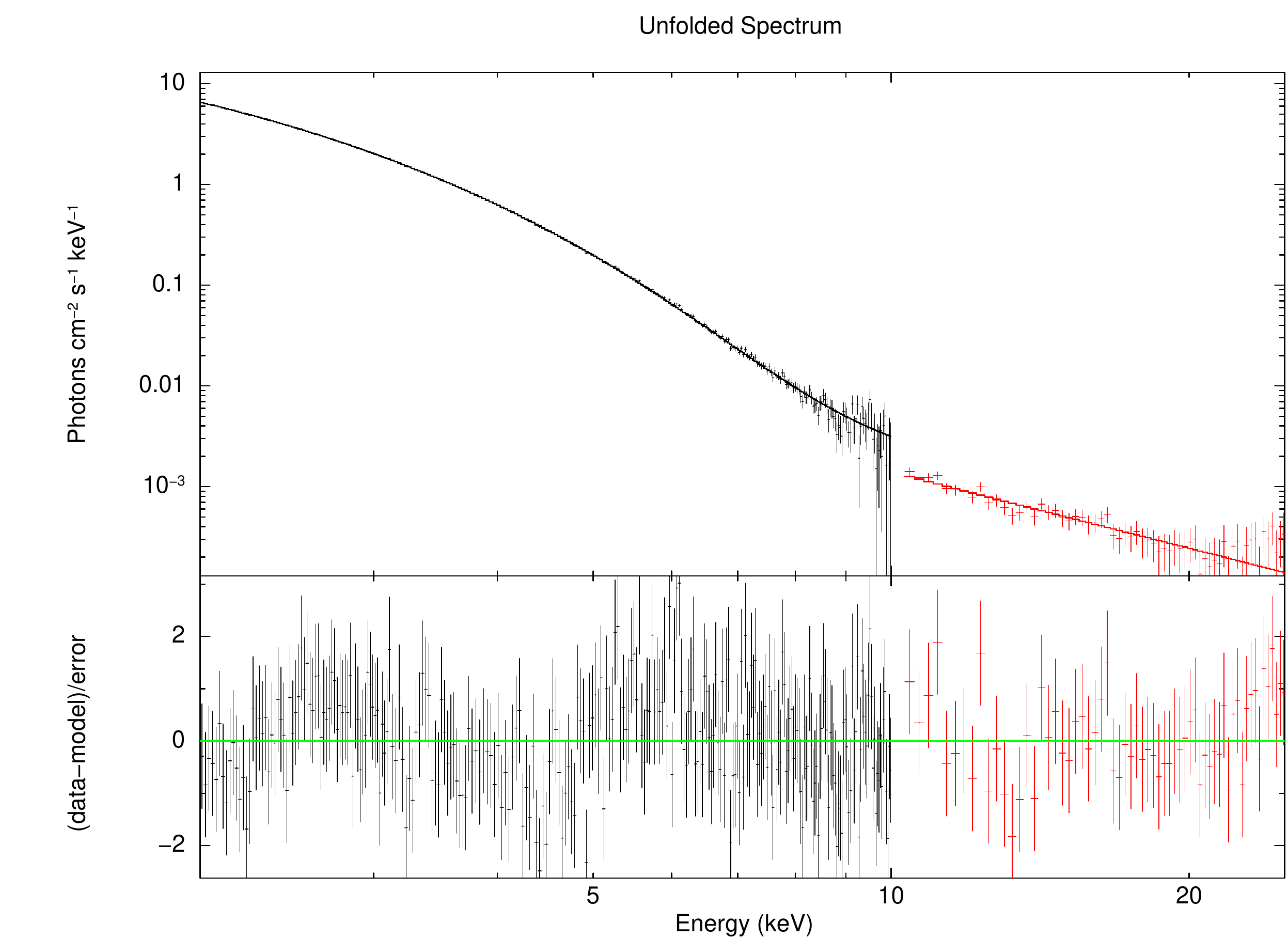}
\caption{A representative (ObsID P011466110701) unfolded spectrum. The spectrum was rebinned just for the purpose of display. \label{fig:rep}}
\end{figure}

For all the``gold" spectra, our model provides a good fit, with the reduced $\chi^2_\nu$ varying between 0.88-1.35. A representative plot of the unfolded spectrum is given in Figure~\ref{fig:rep}. The best-fit value of $a_*$ lies in the range from 0.016 to 0.253, with the mean of 0.153, confirming a slowly-spin black hole in MAXI J1820+070. 

As we mentioned earlier, in order to have a  successful application of the continuum-fitting method, it is important to ensure a disk with the bolometric Eddington-scaled luminosity $l < 0.3$. For MAXI J1820+070, it is clear that our selected ``gold" spectra satisfy this standard, with $l$ ranging from 0.056 to 0.150.

\subsection{Error Analysis}
The errors quoted in Table~\ref{diskbb} and Table~\ref{kerrbb2} are only due to the statistic uncertainties estimated via {\sc Xspec}. The confidence level is 90\%. As mentioned in Section~\ref{into}, the error budget of $a_*$ is dominated by the combined observational uncertainties of $M$, $i$ and $D$. Herein the Monte Carlo (MC) simulation was performed for error analysis. The steps of error analysis are described as following: for each individual spectrum, (1) assuming independent and Gaussian distributed\footnote{Previous work has confirmed that if we generate parameter sets via the mass function, the results will be roughly the same and the difference between these two methods will be nearly negligible (\citealt{Zhao2020}).}, we generate 3000 sets of ($M$, $i$, $D$). (2) we calculate look-up tables of $f$ for these parameter sets. (3) we re-fit the spectrum 3000 times with these ($M$, $i$, $D$) to determine the histograms of $a_{*}$, from which we finally decide the error of $a_*$. 

\begin{figure}[ht!]
\plotone{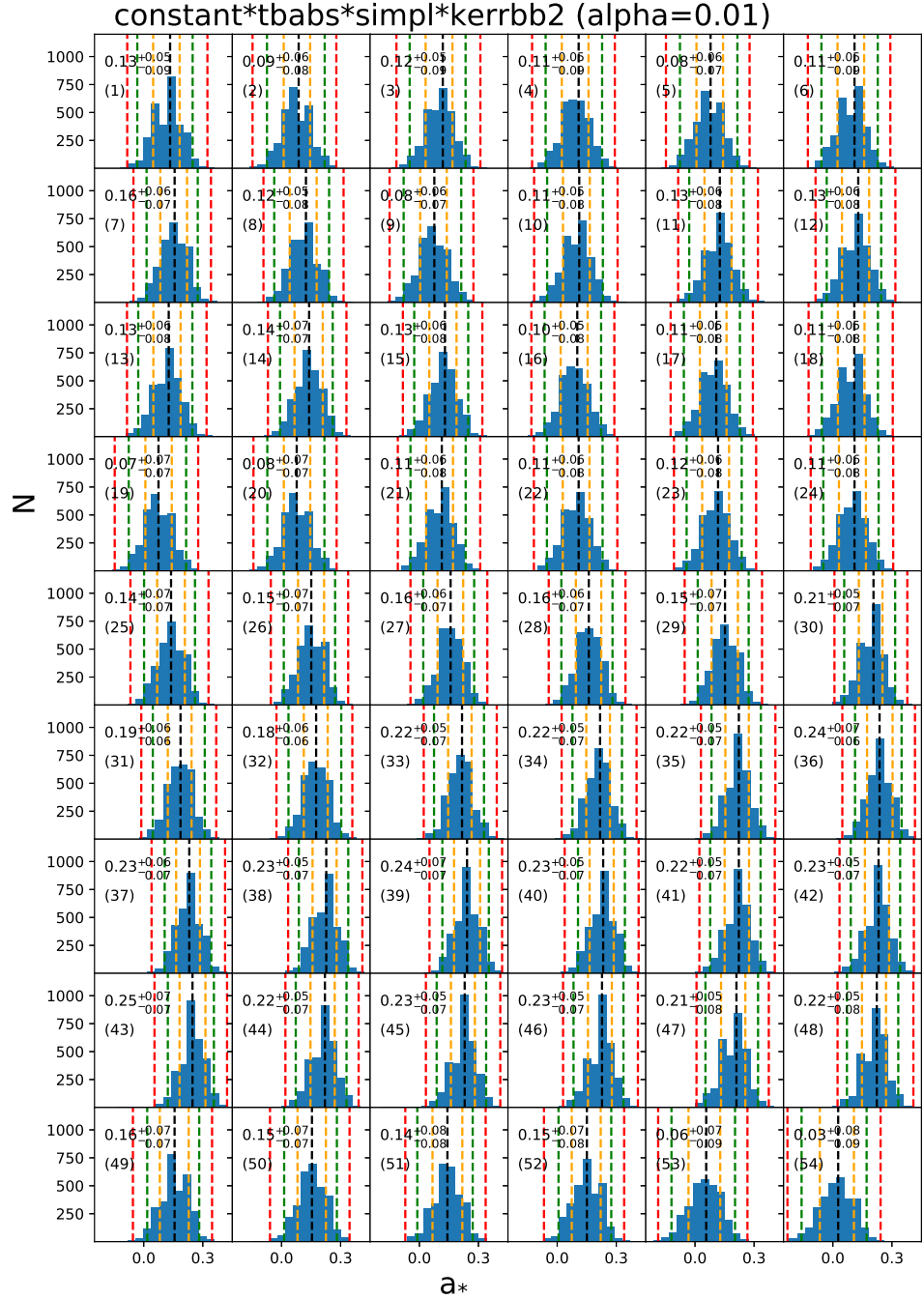}
\caption{Histograms of $a_{*}$ calculated via the Monte Carlo analysis for 3000 sets of parameters per spectrum. The three dashed lines imply the 99.7\% (3$\sigma$, red), 95.4\% (2$\sigma$, green), and 68.3\% (1$\sigma$, orange) error, respectively. The respective 68.3\% confidence level on $a_*$ is indicated in each panel. \label{fig:simpl}}
\end{figure}

\begin{figure}[ht!]
\plotone{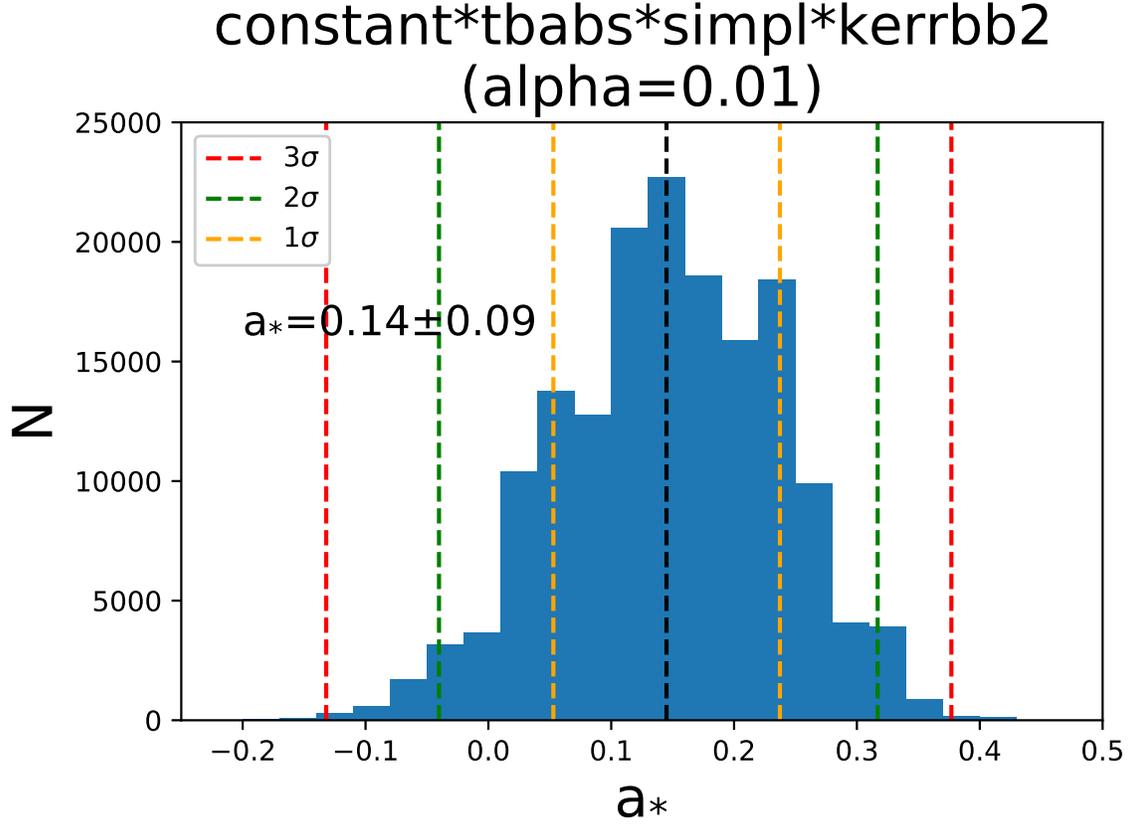}
\caption{Summed histogram of $a_{*}$ for 54 spectra, including 162000 data points.\label{fig:summed}}
\end{figure}

We made MC simulations on 54 ``gold" spectra. Histograms of $a_*$ are plotted in Figure ~\ref{fig:simpl}, and the summed histogram is showed in Figure ~\ref{fig:summed}. Adopting the histogram, we arrived at the final value of $a_*=0.14 \pm 0.09$ ($1\sigma$).

\section{discussion}
\label{discussion}

\subsection{effect of varying $\Gamma$ and normalization}

Normally we expect that if we fix the normalization of LE to 1, and then the normalization of ME will also be 1. However, due to the effects of systematic errors, there are minor differences between the calibration of the two detectors (\citealt{Li2020}), and during fitting process, the relative differences may change slightly (empirically, as for \emph{Insight}-HXMT, within the range of 0.85 to 1.15 is considered as reasonable). We have tried to let the normalization vary between 0.95 to 1.05, however it always peg at its lower limitation 0.95 probably due to the weak Comptonization component, so that we fix the normalization of ME to 0.95.

We tested the effect of different $\Gamma$ and norm on $a_*$, and fit results are presented in Table~\ref{gamma}. It is showed that $a_*$ decreases by $\triangle a_*=0.053$ as $\Gamma$ increases from 2.10 to 2.90 (with norm frozen at 0.95), and increases by $\triangle a_*=0.017$ as norm varies from 0.95 to 1.05 (with $\Gamma$ fix at 2.50).

\subsection{effect of different parameter configurations}

\begin{figure}[ht!]
\plotone{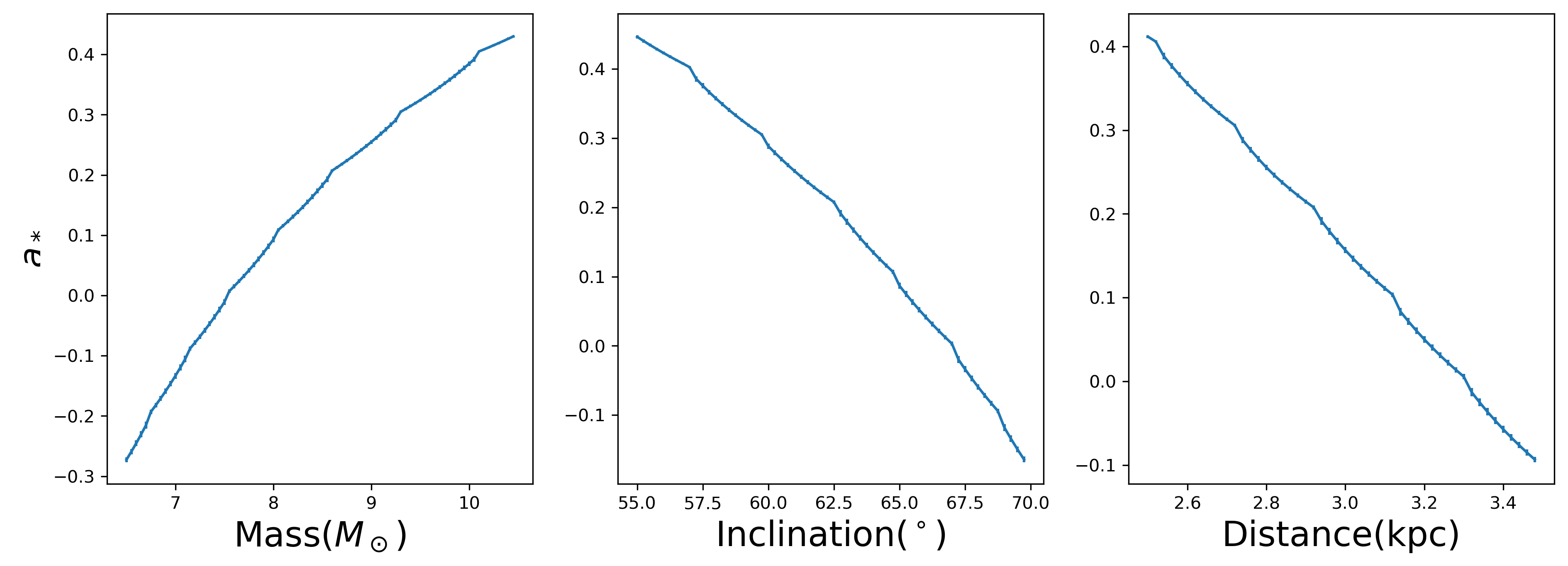}
\caption{The correlation plots displaying the effect on the spin of varying $M$, $i$ and $D$.}
\label{fig:correlation}
\end{figure}

As we mentioned above, the inclination is a crucial input parameter and has an obvious degeneracy with the spin. In previous section, we have the estimated the spin by assuming the jet inclination angle $i=63^\circ\pm3^\circ$, and the black hole mass $M$ = $8.48^{+0.79}_{-0.72}~M_\odot$. We also constrained the spin for the case that the inclination angle ranges between $66\fdg2<i<80\fdg8$. Referring to Figure~\ref{fig:correlation}, the spin parameter decreases as the mass decreases or the inclination increases (See Section~\ref{into} for qualitative analysis), therefore we used $i=66\fdg2$ and $M=8.06~M_\odot$ to constrain the upper limit of the spin for Torres's parameters. The fit results are shown in Table~\ref{torres}. The best-fit values of this set of parameters (within a range of -0.276 to 0.036 and only eleven spectra have a spin larger than zero) are lower than that of our adopted parameters, so that Torres's parameter configuration may leads to a retrograde black hole, which needs to be checked with the precise system parameters in the future. In addition, assuming $M=8.06~M_\odot$, the spin will peg to its lower limit -0.99 in {\sc kerrbb2} if the inclination is above $76^\circ$. The critical value of $i$, which will lead to a spin of -0.99, is $67^\circ$ for $M=5.96~M_\odot$. 

In applying the continuum-fitting method, the inclination $i$ is supposed to be the inclination angle of the inner disk, which, however, is hard to be estimated in practice (In the future, the X-ray polarization method may provide more accurate constraints on it). Usually the strategy is to use the orbital inclination or jet inclination as the proxy, instead. Some previous work on fitting the reflection component reported a small misalignment between the inner accretion disk and the binary orbital plane (\citealt{Fragos2010}, \citealt{Walton2016}), which can be interpreted by a warp in the disk. In this work, we preferred the spin result for adopting the jet inclination angle as the inner disk inclination, however, more consistent and accurate dynamical parameters are required for the detailed spin measurements in the future.

As a caveat, it is noted that all the spin measurements from the continuum-fitting method (\citealt{Liu2008}, \citealt{Gou2009}, \citealt{Gou2010}, \citealt{Steiner2011}, \citealt{Steiner2012},\citealt{Gou2014}, \citealt{Steiner2014}, \citealt{Chen2016}, \citealt{Steiner2016}) basically suggested positive spin values, and there has been no clear observational evidence for the existence of retrograde black holes. \cite{Morningstar2014} initially reported a retrograde spin for the black hole in Nova Muscae 1991: $a_* = -0.25^{+0.05}_{-0.64}$ (90\% confidence level). However, \cite{Chen2016} found a moderately high value of spin, $a_* = 0.63^{+0.16}_{-0.19}$ (1$\sigma$ confidence level) after the system parameters were updated consistently, and rule strongly against a retrograde value: $a_* > 0.17$ (2$\sigma$ or 95.4\% confidence level).





\section{Conclusion}

In this work, we have presented a methodology of the spin measurements for the newly-observed black hole X-ray binary MAXI J1820+070 using \emph{Insight}-HXMT spectra. Mainly because the spin of the black hole strongly depends on the measurement of the disk inclination, the black hole mass and the distance, the large uncertainty of the dynamical parameters makes it difficult to critically evaluate its spin. For MAXI J1820+070, we have discussed two scenarios. Preferring to consider the jet inclination as the inner accretion disk inclination angle, adopting $M$ = $8.48^{+0.79}_{-0.72}~M_\odot$, $i=63^\circ\pm3^\circ$ and $D=2.96\pm0.33$ kpc, we deduce a value of $a_*=0.14 \pm 0.09$ ($1\sigma$), showing that the black hole in this system is rotating slowly. Besides, when the parameter ranges $5.96~M_\odot < M < 8.06~M_\odot$ and $66\fdg2< i < 80\fdg8$ are applied, the black hole are more likely to have a retrograde spin. 

\cite{Shidatsu2019} estimated the inner edge of the accretion disk in the HSS to be $R_{\rm in}$ = 77.9$\pm$1.0($D$/3 kpc)(cos$i$/cos60$^{\circ}$)$^{-1/2}$~km. Using $M$ = $8.48^{+0.79}_{-0.72}~M_\odot$, $i=63^\circ\pm3^\circ$ and $D=2.96\pm0.33$ kpc, the black hole in MAXI J1820+070 is near non-rotating, which is basically compatible with our constrain of $a_*=0.14 \pm 0.09$ ($1\sigma$). In addition, it is noted that the reflection spectral fit to the NuSTAR prefers a smaller inner disk inclination between 30$^\circ$-40$^\circ$  and the spin range between -0.5 and 0.5 (\citealt{Bharali2019}, \citealt{Buisson2019}, \citealt{Fabian2020}). However,  if the disk inclination is set around $70^\circ$, the spin derived from the same soft-state NuSTAR spectra would be almost maximally retrograde ($a_* < -0.95$, \citealt{Fabian2020}). In any event, we hope future observations could help improve the dynamical parameters, hence putting a tighter constraints on the spin parameters.

\begin{deluxetable*}{lllrrrrrrll}
\tablecaption{\emph{Insight}-HXMT Observational Journal of MAXI J1820+070\label{journal}$^a$}
\tablewidth{700pt}
\tabletypesize{\scriptsize}
\tablehead{
\colhead{Number} & \colhead{ObsID} & 
\colhead{MJD} & \colhead{Start Time} & 
\colhead{End Time} & \colhead{exposure(s)}  & \colhead{State$^b$}\\
} 
\startdata
1	&	P011466108401	&	58305.86277 	&	2018-07-06T20:41:17	&	2018-07-06T23:48:23	&	1674 	$\&$	2155 			&	IM	\\
2	&	P011466108402	&	58305.99270 	&	2018-07-06T23:48:23	&	2018-07-07T03:12:37	&	1374 	$\&$	2556 			&	IM	\\
3	&	P011466108403	&	58306.13453 	&	2018-07-07T03:12:37	&	2018-07-07T06:22:11	&	810 	$\&$	1658 			&	IM/HSS	\\
4	&	P011466108501	&	58307.05589 	&	2018-07-08T01:19:23	&	2018-07-08T04:43:34	&	1211 	$\&$	2014 			&	IM/HSS	\\
5	&	P011466108502	&	58307.19769 	&	2018-07-08T04:43:34	&	2018-07-08T09:24:42	&	695 	$\&$	1534 			&	IM/HSS	\\
6	&	P011466108601	&	58308.11641 	&	2018-07-09T02:46:32	&	2018-07-09T06:14:59	&	747 	$\&$	1114 			&	IM/HSS	\\
7	&	P011466108602	&	58308.26117 	&	2018-07-09T06:14:59	&	2018-07-09T10:51:44	&	539 	$\&$	1799 			&	IM/HSS	\\
8	&	P011466108702	&	58309.25513 	&	2018-07-10T06:06:17	&	2018-07-10T09:07:54	&	479 	$\&$	1423 			&	IM/HSS	\\
9	&	P011466108801	&	58310.03858 	&	2018-07-11T00:54:27	&	2018-07-11T04:17:01	&	1556 	$\&$	1343 			&	IM/HSS	\\
10	&	P011466108802	&	58310.17925 	&	2018-07-11T04:17:01	&	2018-07-11T07:24:09	&	898 	$\&$	941 			&	IM/HSS	\\
11	&	P011466108901	&	58311.23160 	&	2018-07-12T05:32:24	&	2018-07-12T10:29:02	&	931 	$\&$	2358 			&	HSS	\\
12	&	P011466109001	&	58312.82233 	&	2018-07-13T19:43:03	&	2018-07-14T00:44:39	&	2134 	$\&$	2934 			&	HSS	\\
13	&	P011466109101	&	58314.08167 	&	2018-07-15T01:56:30	&	2018-07-15T05:23:05	&	1002 	$\&$	1027 			&	HSS	\\
14	&	P011466109102	&	58314.22513 	&	2018-07-15T05:23:05	&	2018-07-15T09:07:27	&	1195 	$\&$	1968 			&	HSS	\\
15	&	P011466109201	&	58316.79946 	&	2018-07-17T19:10:07	&	2018-07-17T22:15:11	&	1077 	$\&$	1903 			&	HSS	\\
16	&	P011466109202	&	58316.92798 	&	2018-07-17T22:15:11	&	2018-07-18T01:39:17	&	1305 	$\&$	1272 			&	HSS	\\
17	&	P011466109301	&	58317.66130 	&	2018-07-18T15:51:10	&	2018-07-18T20:44:08	&	1077 	$\&$	3420 			&	HSS	\\
18	&	P011466109401	&	58318.45692 	&	2018-07-19T10:56:52	&	2018-07-19T15:52:51	&	1257 	$\&$	3160 			&	HSS	\\
19	&	P011466109501	&	58320.71167 	&	2018-07-21T17:03:42	&	2018-07-21T20:59:14	&	3039 	$\&$	2818 			&	HSS	\\
20	&	P011466109503	&	58321.00778 	&	2018-07-22T00:10:06	&	2018-07-22T02:50:58	&	479 	$\&$	760 			&	HSS	\\
21	&	P011466109601	&	58322.76842 	&	2018-07-23T18:25:25	&	2018-07-23T22:17:50	&	2298 	$\&$	2554 			&	HSS	\\
22	&	P011466109602	&	58322.92982 	&	2018-07-23T22:17:50	&	2018-07-24T02:28:17	&	1270 	$\&$	1777 			&	HSS	\\
23	&	P011466109701	&	58324.95915 	&	2018-07-25T23:00:04	&	2018-07-26T02:47:29	&	1564 	$\&$	2244 			&	HSS	\\
24	&	P011466109702	&	58325.11707 	&	2018-07-26T02:47:29	&	2018-07-26T05:27:31	&	1583 	$\&$	1776 			&	HSS	\\
25	&	P011466109801	&	58326.22091 	&	2018-07-27T05:17:00	&	2018-07-27T09:01:17	&	2454 	$\&$	2360 			&	HSS	\\
26	&	P011466109802	&	58326.37666 	&	2018-07-27T09:01:17	&	2018-07-27T12:12:12	&	2453 	$\&$	2228 			&	HSS	\\
27	&	P011466109803	&	58326.50924 	&	2018-07-27T12:12:12	&	2018-07-27T14:55:38	&	2213 	$\&$	1981 			&	HSS	\\
28	&	P011466109901	&	58327.08413 	&	2018-07-28T02:00:03	&	2018-07-28T06:50:58	&	2544 	$\&$	3874 			&	HSS	\\
29	&	P011466110001	&	58328.94229 	&	2018-07-29T22:35:48	&	2018-07-30T03:26:25	&	2523 	$\&$	4027 			&	HSS	\\
30	&	P011466110101	&	58330.60005 	&	2018-07-31T14:22:58	&	2018-07-31T19:13:23	&	2261 	$\&$	4219 			&	HSS	\\
31	&	P011466110201	&	58331.66060 	&	2018-08-01T15:50:10	&	2018-08-01T20:40:32	&	3700 	$\&$	3544 			&	HSS	\\
32	&	P011466110301	&	58332.45593 	&	2018-08-02T10:55:26	&	2018-08-02T14:35:40	&	2101 	$\&$	4263 			&	HSS	\\
33	&	P011466110302	&	58332.60887 	&	2018-08-02T14:35:40	&	2018-08-02T18:56:37	&	2573 	$\&$	3701 			&	HSS	\\
34	&	P011466110401	&	58334.51027 	&	2018-08-04T12:13:41	&	2018-08-04T15:55:25	&	2829 	$\&$	4762 			&	HSS	\\
35	&	P011466110701	&	58335.50424 	&	2018-08-05T12:05:00	&	2018-08-05T16:55:15	&	4219 	$\&$	5276 			&	HSS	\\
36	&	P011466110801	&	58336.29941 	&	2018-08-06T07:10:03	&	2018-08-06T12:00:50	&	1841 	$\&$	3871 			&	HSS	\\
37	&	P011466110901	&	58338.55235 	&	2018-08-08T13:14:17	&	2018-08-08T16:58:34	&	3286 	$\&$	4320 			&	HSS	\\
38	&	P011466110902	&	58338.70810 	&	2018-08-08T16:58:34	&	2018-08-08T19:40:25	&	1137 	$\&$	1704 			&	HSS	\\
39	&	P011466111001	&	58339.48006 	&	2018-08-09T11:30:11	&	2018-08-09T16:20:53	&	3044 	$\&$	4386 			&	HSS	\\
40	&	P011466111201	&	58341.60057 	&	2018-08-11T14:23:43	&	2018-08-11T19:20:41	&	2514 	$\&$	3491 			&	HSS	\\
41	&	P011466111301	&	58343.52242 	&	2018-08-13T12:31:11	&	2018-08-13T17:21:47	&	2155 	$\&$	3185 			&	HSS	\\
42	&	P011466111401	&	58345.24557 	&	2018-08-15T05:52:31	&	2018-08-15T10:43:08	&	2692 	$\&$	3519 			&	HSS	\\
43	&	P011466111501	&	58347.23405 	&	2018-08-17T05:35:56	&	2018-08-17T10:26:31	&	2514 	$\&$	3516 			&	HSS	\\
44	&	P011466111601	&	58349.68682 	&	2018-08-19T16:27:55	&	2018-08-19T21:18:31	&	1735 	$\&$	2693 			&	HSS	\\
45	&	P011466111701	&	58352.47157 	&	2018-08-22T11:17:57	&	2018-08-22T16:08:32	&	1648 	$\&$	2758 			&	HSS	\\
46	&	P011466111801	&	58356.71551 	&	2018-08-26T17:09:14	&	2018-08-26T20:09:34	&	599 	$\&$	700 			&	HSS	\\
47	&	P011466111802	&	58356.84074 	&	2018-08-26T20:09:34	&	2018-08-27T01:15:46	&	180 	$\&$	658 			&	HSS	\\
48	&	P011466111901	&	58358.50582 	&	2018-08-28T12:07:17	&	2018-08-28T16:57:46	&	1542 	$\&$	3007 			&	HSS	\\
49	&	P011466112001	&	58359.89813 	&	2018-08-29T21:32:12	&	2018-08-30T02:23:11	&	1317 	$\&$	1938 			&	HSS	\\
50	&	P011466112101	&	58361.95323 	&	2018-08-31T22:51:33	&	2018-09-01T03:42:27	&	1594 	$\&$	3417 			&	HSS	\\
51	&	P011466112201	&	58364.60463 	&	2018-09-03T14:29:34	&	2018-09-03T19:20:11	&	958 	$\&$	1976 			&	HSS	\\
52	&	P011466112301	&	58366.92445 	&	2018-09-05T22:10:06	&	2018-09-06T01:27:46	&	406 	$\&$	2095 			&	HSS	\\
53	&	P011466112401	&	58376.07197 	&	2018-09-15T01:42:32	&	2018-09-15T05:26:46	&	1834 	$\&$	2182 			&	HSS	\\
54	&	P011466112402	&	58376.22769 	&	2018-09-15T05:26:46	&	2018-09-15T08:08:25	&	539 	$\&$	941 			&	HSS	\\
55	&	P011466112501	&	58377.46444 	&	2018-09-16T11:07:41	&	2018-09-16T15:58:08	&	3111 	$\&$	2702 			&	HSS	\\
56	&	P011466112601	&	58379.32135 	&	2018-09-18T07:41:38	&	2018-09-18T12:35:16	&	2655 	$\&$	2825 			&	HSS	\\
57	&	P011466112701	&	58381.04584 	&	2018-09-20T01:04:54	&	2018-09-20T05:55:25	&	2076 	$\&$	2251 			&	IM	\\
58	&	P011466112801	&	58382.63775 	&	2018-09-21T15:17:15	&	2018-09-21T20:10:31	&	1900 	$\&$	1990 			&	IM	\\
59	&	P011466112901	&	58383.43367 	&	2018-09-22T10:23:23	&	2018-09-22T15:16:09	&	3638 	$\&$	3029 			&	IM	\\
60	&	P011466113001	&	58385.15798 	&	2018-09-24T03:46:23	&	2018-09-24T08:36:49	&	2910 	$\&$	2894 			&	IM	\\
61	&	P011466113101	&	58387.14719 	&	2018-09-26T03:30:51	&	2018-09-26T08:21:12	&	3694 	$\&$	3042 			&	IM	\\
\enddata
\tablecomments{$^a$The log of \emph{Insight}-HXMT observations analyzed in this work. In columns 2–7, we show the following information, respectively: observation ID; MJD; the start time of the observations; the end time of the observations; the effective exposure times in units of second for LE and ME;  accretion states. \\
$^b$We adopted the states defined in \cite{Shidatsu2019}. "IM" and "HSS" represent the intermediate and high/soft state, respectively. It is noted that \cite{Buisson2019} classified spectra 3-10 as HSS.\\
}
\end{deluxetable*}

\begin{deluxetable*}{lllrrrrrrll}
\tablecaption{Best-fit parameters for spectra with the model {\sc constant*tbabs*simpl*diskbb} \\($M$ = $8.48~M_\odot$, $i=63^\circ$ and $D=2.96$ kpc)\label{diskbb}}
\tablewidth{700pt}
\tabletypesize{\scriptsize}
\tablehead{
\colhead{Number} & \colhead{ObsID} & \multicolumn{2}{c}{\sc simpl} & \multicolumn{2}{c}{\sc diskbb}  & \colhead{Reduced $\chi^2_\nu$} & \colhead{$\chi^2$/d.o.f.}\\
\cline{3-6}
\colhead{} & \colhead{} &\colhead{$\Gamma$} & \colhead{$f_{\rm sc}$} & \colhead{$T_{\rm in}$} & \colhead{$R_{\rm in}$} & \colhead{norm}
} 
\startdata
1&P011466108502 & 2.36 $\pm$ 0.02 & 0.067 $\pm$ 0.001 & 0.748 $\pm$ 0.002 & 5.28 $\pm$ 0.58 & 1.64 & 1460.3 / 893 \\
2&P011466108601 & 2.48 $\pm$ 0.02 & 0.064 $\pm$ 0.002 & 0.743 $\pm$ 0.002 & 5.49 $\pm$ 0.60 & 1.47 & 1313.5 / 893 \\
3&P011466108602 & 2.23 $\pm$ 0.01 & 0.059 $\pm$ 0.001 & 0.752 $\pm$ 0.002 & 5.29 $\pm$ 0.58 & 1.26 & 1085.9 / 862 \\
4&P011466108702 & 2.27 $\pm$ 0.02 & 0.036 $\pm$ 0.001 & 0.753 $\pm$ 0.002 & 5.36 $\pm$ 0.58 & 1.35 & 1098.7 / 816 \\
5&P011466108801 & 2.41 $\pm$ 0.02 & 0.037 $\pm$ 0.001 & 0.743 $\pm$ 0.001 & 5.53 $\pm$ 0.51 & 1.43 & 1340.6 / 940 \\
6&P011466108802 & 2.35 $\pm$ 0.03 & 0.033 $\pm$ 0.001 & 0.745 $\pm$ 0.002 & 5.48 $\pm$ 0.55 & 1.33 & 1157.1 / 872 \\
7&P011466108901 & 2.18 $\pm$ 0.03 & 0.017 $\pm$ 0.001 & 0.758 $\pm$ 0.001 & 5.16 $\pm$ 0.47 & 1.29 & 1103.0 / 855 \\
8&P011466109001 & 2.31 $\pm$ 0.02 & 0.024 $\pm$ 0.001 & 0.751 $\pm$ 0.001 & 5.33 $\pm$ 0.44 & 1.69 & 1611.1 / 952 \\
9&P011466109101 & 2.45 $\pm$ 0.03 & 0.035 $\pm$ 0.001 & 0.738 $\pm$ 0.002 & 5.55 $\pm$ 0.55 & 1.24 & 1086.2 / 876 \\
10&P011466109102 & 2.24 $\pm$ 0.02 & 0.025 $\pm$ 0.001 & 0.748 $\pm$ 0.001 & 5.37 $\pm$ 0.48 & 1.40 & 1247.7 / 893 \\
11&P011466109201 & 2.26 $\pm$ 0.02 & 0.023 $\pm$ 0.001 & 0.748 $\pm$ 0.001 & 5.31 $\pm$ 0.49 & 1.42 & 1243.1 / 875 \\
12&P011466109202 & 2.43 $\pm$ 0.03 & 0.028 $\pm$ 0.001 & 0.743 $\pm$ 0.001 & 5.39 $\pm$ 0.51 & 1.28 & 1137.8 / 891 \\
13&P011466109301 & 2.16 $\pm$ 0.03 & 0.016 $\pm$ 0.001 & 0.751 $\pm$ 0.001 & 5.25 $\pm$ 0.46 & 1.44 & 1244.8 / 863 \\
14&P011466109401 & 2.20 $\pm$ 0.03 & 0.014 $\pm$ 0.001 & 0.750 $\pm$ 0.001 & 5.21 $\pm$ 0.45 & 1.46 & 1274.5 / 870 \\
15&P011466109501 & 2.56 $\pm$ 0.03 & 0.023 $\pm$ 0.001 & 0.741 $\pm$ 0.001 & 5.35 $\pm$ 0.43 & 1.56 & 1498.3 / 963 \\
16&P011466109503 & 2.42 $\pm$ 0.04 & 0.026 $\pm$ 0.002 & 0.736 $\pm$ 0.002 & 5.46 $\pm$ 0.62 & 1.09 & 791.6 / 724 \\
17&P011466109601 & 2.38 $\pm$ 0.02 & 0.027 $\pm$ 0.001 & 0.735 $\pm$ 0.001 & 5.44 $\pm$ 0.46 & 1.49 & 1420.8 / 952 \\
18&P011466109602 & 2.34 $\pm$ 0.02 & 0.027 $\pm$ 0.001 & 0.734 $\pm$ 0.001 & 5.44 $\pm$ 0.51 & 1.43 & 1272.2 / 889 \\
19&P011466109701 & 2.40 $\pm$ 0.02 & 0.030 $\pm$ 0.001 & 0.728 $\pm$ 0.001 & 5.51 $\pm$ 0.50 & 1.50 & 1375.2 / 918 \\
20&P011466109702 & 2.41 $\pm$ 0.02 & 0.029 $\pm$ 0.001 & 0.727 $\pm$ 0.001 & 5.50 $\pm$ 0.50 & 1.53 & 1391.2 / 912 \\
21&P011466109801 & 2.57 $\pm$ 0.03 & 0.023 $\pm$ 0.001 & 0.727 $\pm$ 0.001 & 5.43 $\pm$ 0.46 & 1.42 & 1322.1 / 932 \\
22&P011466109802 & 2.63 $\pm$ 0.03 & 0.026 $\pm$ 0.001 & 0.724 $\pm$ 0.001 & 5.47 $\pm$ 0.47 & 1.51 & 1410.5 / 932 \\
23&P011466109803 & 2.77 $\pm$ 0.03 & 0.028 $\pm$ 0.001 & 0.724 $\pm$ 0.001 & 5.47 $\pm$ 0.49 & 1.40 & 1283.1 / 919 \\
24&P011466109901 & 2.66 $\pm$ 0.03 & 0.019 $\pm$ 0.001 & 0.726 $\pm$ 0.001 & 5.42 $\pm$ 0.45 & 1.40 & 1289.8 / 924 \\
25&P011466110001 & 4.05 $\pm$ 0.07 & 0.047 $\pm$ 0.003 & 0.713 $\pm$ 0.002 & 5.52 $\pm$ 0.55 & 1.19 & 1077.8 / 902 \\
26&P011466110101 & 3.78 $\pm$ 0.08 & 0.032 $\pm$ 0.003 & 0.718 $\pm$ 0.001 & 5.37 $\pm$ 0.52 & 1.22 & 1076.5 / 885 \\
27&P011466110201 & 4.50 $\pm$ 0.07 & 0.075 $\pm$ 0.005 & 0.699 $\pm$ 0.002 & 5.60 $\pm$ 0.57 & 1.19 & 1110.8 / 932 \\
28&P011466110301 & 4.50 $\pm$ 0.09 & 0.066 $\pm$ 0.005 & 0.700 $\pm$ 0.002 & 5.55 $\pm$ 0.61 & 1.14 & 997.6 / 875 \\
29&P011466110302 & 4.50 $\pm$ 0.09 & 0.066 $\pm$ 0.005 & 0.698 $\pm$ 0.002 & 5.59 $\pm$ 0.60 & 1.12 & 1000.0 / 893 \\
30&P011466110401 & 4.50 $\pm$ 0.07 & 0.078 $\pm$ 0.005 & 0.697 $\pm$ 0.002 & 5.52 $\pm$ 0.59 & 1.15 & 1042.1 / 903 \\
31&P011466110701 & 4.50 $\pm$ 0.07 & 0.071 $\pm$ 0.004 & 0.694 $\pm$ 0.002 & 5.52 $\pm$ 0.55 & 1.25 & 1175.5 / 941 \\
32&P011466110801 & 4.50 $\pm$ 0.09 & 0.068 $\pm$ 0.006 & 0.691 $\pm$ 0.002 & 5.55 $\pm$ 0.64 & 1.19 & 1017.9 / 854 \\
33&P011466110901 & 5.00 $\pm$ 0.17 & 0.105 $\pm$ 0.008 & 0.680 $\pm$ 0.002 & 5.60 $\pm$ 0.64 & 1.03 & 915.7 / 891 \\
34&P011466110902 & 4.50 $\pm$ 0.11 & 0.077 $\pm$ 0.008 & 0.687 $\pm$ 0.003 & 5.47 $\pm$ 0.71 & 0.91 & 707.6 / 780 \\
35&P011466111001 & 4.50 $\pm$ 0.05 & 0.109 $\pm$ 0.006 & 0.675 $\pm$ 0.002 & 5.65 $\pm$ 0.63 & 1.61 & 1454.8 / 902 \\
36&P011466111201 & 4.50 $\pm$ 0.07 & 0.098 $\pm$ 0.006 & 0.675 $\pm$ 0.002 & 5.50 $\pm$ 0.63 & 1.46 & 1274.3 / 871 \\
37&P011466111301 & 5.00 $\pm$ 0.21 & 0.104 $\pm$ 0.009 & 0.672 $\pm$ 0.003 & 5.53 $\pm$ 0.70 & 0.91 & 798.0 / 875 \\
38&P011466111401 & 4.50 $\pm$ 0.10 & 0.065 $\pm$ 0.006 & 0.672 $\pm$ 0.002 & 5.44 $\pm$ 0.61 & 1.07 & 921.6 / 865 \\
39&P011466111501 & 4.50 $\pm$ 0.07 & 0.090 $\pm$ 0.006 & 0.662 $\pm$ 0.002 & 5.52 $\pm$ 0.65 & 1.36 & 1176.2 / 862 \\
40&P011466111601 & 4.50 $\pm$ 0.07 & 0.104 $\pm$ 0.007 & 0.648 $\pm$ 0.002 & 5.66 $\pm$ 0.73 & 1.33 & 1078.6 / 813 \\
41&P011466111701 & 4.50 $\pm$ 0.10 & 0.080 $\pm$ 0.007 & 0.647 $\pm$ 0.002 & 5.58 $\pm$ 0.73 & 0.98 & 781.9 / 798 \\
42&P011466111801 & 4.15 $\pm$ 0.12 & 0.084 $\pm$ 0.010 & 0.637 $\pm$ 0.004 & 5.65 $\pm$ 0.92 & 1.02 & 547.1 / 537 \\
43&P011466111802 & 4.50 $\pm$ 0.14 & 0.147 $\pm$ 0.023 & 0.614 $\pm$ 0.008 & 6.04 $\pm$ 1.43 & 1.05 & 446.7 / 424 \\
44&P011466111901 & 4.50 $\pm$ 0.09 & 0.095 $\pm$ 0.008 & 0.623 $\pm$ 0.003 & 5.81 $\pm$ 0.87 & 0.81 & 637.0 / 783 \\
45&P011466112001 & 4.50 $\pm$ 0.11 & 0.092 $\pm$ 0.009 & 0.619 $\pm$ 0.003 & 5.79 $\pm$ 0.85 & 1.00 & 742.9 / 742 \\
46&P011466112101 & 4.50 $\pm$ 0.09 & 0.091 $\pm$ 0.008 & 0.614 $\pm$ 0.003 & 5.79 $\pm$ 0.81 & 0.86 & 661.4 / 772 \\
47&P011466112201 & 4.31 $\pm$ 0.08 & 0.114 $\pm$ 0.009 & 0.591 $\pm$ 0.003 & 6.21 $\pm$ 1.01 & 1.03 & 733.1 / 713 \\
48&P011466112301 & 4.50 $\pm$ 0.10 & 0.136 $\pm$ 0.015 & 0.580 $\pm$ 0.005 & 6.36 $\pm$ 1.30 & 0.96 & 592.2 / 616 \\
49&P011466112401 & 4.18 $\pm$ 0.09 & 0.085 $\pm$ 0.007 & 0.563 $\pm$ 0.003 & 6.25 $\pm$ 0.94 & 0.93 & 688.3 / 743 \\
50&P011466112402 & 3.60 $\pm$ 0.12 & 0.049 $\pm$ 0.006 & 0.574 $\pm$ 0.004 & 5.94 $\pm$ 1.05 & 0.89 & 419.1 / 473 \\
51&P011466112501 & 3.65 $\pm$ 0.06 & 0.056 $\pm$ 0.004 & 0.567 $\pm$ 0.002 & 6.04 $\pm$ 0.76 & 0.96 & 771.9 / 805 \\
52&P011466112601 & 3.53 $\pm$ 0.06 & 0.047 $\pm$ 0.003 & 0.564 $\pm$ 0.002 & 5.98 $\pm$ 0.76 & 0.96 & 741.8 / 772 \\
53&P011466112701 & 2.85 $\pm$ 0.04 & 0.030 $\pm$ 0.002 & 0.557 $\pm$ 0.002 & 6.00 $\pm$ 0.76 & 0.99 & 738.1 / 746 \\
54&P011466112801 & 2.43 $\pm$ 0.03 & 0.033 $\pm$ 0.001 & 0.547 $\pm$ 0.002 & 6.11 $\pm$ 0.80 & 0.97 & 735.8 / 761 \\
\hline
55&P011466108401 & 2.51 $\pm$ 0.01 & 0.186 $\pm$ 0.002 & 0.736 $\pm$ 0.002 & 5.38 $\pm$ 0.60 & 2.05 & 2151.0 / 1051 \\
56&P011466108402 & 2.46 $\pm$ 0.01 & 0.174 $\pm$ 0.002 & 0.751 $\pm$ 0.002 & 5.16 $\pm$ 0.58 & 2.42 & 2512.2 / 1037 \\
57&P011466108403 & 2.49 $\pm$ 0.01 & 0.174 $\pm$ 0.002 & 0.743 $\pm$ 0.003 & 5.32 $\pm$ 0.66 & 1.90 & 1879.9 / 992 \\
58&P011466108501 & 2.46 $\pm$ 0.01 & 0.102 $\pm$ 0.002 & 0.730 $\pm$ 0.002 & 5.38 $\pm$ 0.58 & 2.10 & 2049.9 / 975 \\
59&P011466112901 & 2.41 $\pm$ 0.01 & 0.100 $\pm$ 0.001 & 0.538 $\pm$ 0.002 & 6.02 $\pm$ 0.80 & 1.25 & 1229.2 / 981 \\
60&P011466113001 & 2.33 $\pm$ 0.01 & 0.119 $\pm$ 0.001 & 0.525 $\pm$ 0.002 & 5.84 $\pm$ 0.88 & 1.22 & 1173.5 / 960 \\
61&P011466113101 & 2.19 $\pm$ 0.01 & 0.182 $\pm$ 0.002 & 0.478 $\pm$ 0.003 & 5.91 $\pm$ 1.09 & 1.09 & 1105.3 / 1014 \\
\enddata
\tablecomments{In column 3-8, we show the following information: the dimensionless photon index of power-law $\Gamma$; the scattered fraction$f_{\rm sc}$; the temperature of inner disk radius $T_{\rm in}$ in the unit of keV; the inner radius $R_{\rm in}$ of the thin accretion disk in the unit of gravitational radius; the reduced chi-square; the total chi-square $\chi^2$ and degrees of freedom d.o.f..}
\end{deluxetable*}

\begin{deluxetable*}{lllrrrrrrll}
\tablecaption{Best-fit parameters for spectra with the model {\sc constant*tbabs*simpl*kerrbb2} \\($\alpha=0.01$, $M$ = $8.48~M_\odot$, $i=63^\circ$ and $D=2.96$ kpc)\label{kerrbb2}}
\tablewidth{700pt}
\tabletypesize{\scriptsize}
\tablehead{
\colhead{Number} & \colhead{ObsID} & \multicolumn{2}{c}{\sc simpl} & \multicolumn{2}{c}{\sc kerrbb2}  & \colhead{Reduced $\chi^2_\nu$} & \colhead{$\chi^2$/d.o.f.} & \colhead{$l^c$}\\
\cline{3-6}
\colhead{} & \colhead{} &\colhead{$\Gamma$} & \colhead{$f_{\rm sc}$} & \colhead{$a_{*}$$^a$} & \colhead{$\dot{M}$$^b$} 
} 
\startdata
1&P011466108502 & 2.24 $\pm$ 0.01 & 0.1016 $\pm$ 0.0022 & 0.129 $\pm$ 0.009 & 2.72 $\pm$ 0.03 & 1.22 & 1130.6 / 923 & 0.142 \\
2&P011466108601 & 2.29 $\pm$ 0.02 & 0.0901 $\pm$ 0.0021 & 0.083 $\pm$ 0.011 & 2.94 $\pm$ 0.03 & 1.18 & 1090.7 / 921 & 0.150 \\
3&P011466108602 & 2.15 $\pm$ 0.01 & 0.0956 $\pm$ 0.0019 & 0.116 $\pm$ 0.009 & 2.81 $\pm$ 0.03 & 1.10 & 974.1 / 887 & 0.146 \\
4&P011466108702 & 2.17 $\pm$ 0.02 & 0.0573 $\pm$ 0.0017 & 0.096 $\pm$ 0.012 & 2.91 $\pm$ 0.03 & 1.10 & 917.5 / 835 & 0.149 \\
5&P011466108801 & 2.15 $\pm$ 0.02 & 0.0472 $\pm$ 0.0012 & 0.076 $\pm$ 0.008 & 2.94 $\pm$ 0.02 & 1.06 & 1037.4 / 980 & 0.149 \\
6&P011466108802 & 2.07 $\pm$ 0.02 & 0.0400 $\pm$ 0.0015 & 0.107 $\pm$ 0.007 & 2.85 $\pm$ 0.02 & 1.02 & 913.7 / 899 & 0.147 \\
7&P011466108901 & 2.04 $\pm$ 0.02 & 0.0258 $\pm$ 0.0010 & 0.155 $\pm$ 0.007 & 2.62 $\pm$ 0.02 & 0.95 & 831.2 / 878 & 0.140 \\
8&P011466109001 & 2.07 $\pm$ 0.02 & 0.0303 $\pm$ 0.0009 & 0.123 $\pm$ 0.005 & 2.75 $\pm$ 0.01 & 1.11 & 1106.3 / 993 & 0.143 \\
9&P011466109101 & 2.18 $\pm$ 0.02 & 0.0437 $\pm$ 0.0015 & 0.072 $\pm$ 0.009 & 2.90 $\pm$ 0.02 & 0.95 & 855.1 / 903 & 0.146 \\
10&P011466109102 & 2.06 $\pm$ 0.02 & 0.0344 $\pm$ 0.0011 & 0.109 $\pm$ 0.005 & 2.78 $\pm$ 0.02 & 0.98 & 907.1 / 923 & 0.144 \\
11&P011466109201 & 2.07 $\pm$ 0.02 & 0.0324 $\pm$ 0.0011 & 0.124 $\pm$ 0.006 & 2.69 $\pm$ 0.02 & 0.99 & 892.5 / 901 & 0.141 \\
12&P011466109202 & 2.08 $\pm$ 0.03 & 0.0304 $\pm$ 0.0012 & 0.127 $\pm$ 0.006 & 2.68 $\pm$ 0.02 & 1.00 & 922.0 / 921 & 0.140 \\
13&P011466109301 & 2.04 $\pm$ 0.02 & 0.0246 $\pm$ 0.0009 & 0.126 $\pm$ 0.006 & 2.67 $\pm$ 0.02 & 1.03 & 911.4 / 888 & 0.139 \\
14&P011466109401 & 2.06 $\pm$ 0.03 & 0.0205 $\pm$ 0.0009 & 0.139 $\pm$ 0.006 & 2.59 $\pm$ 0.02 & 1.01 & 908.1 / 896 & 0.136 \\
15&P011466109501 & 2.14 $\pm$ 0.02 & 0.0236 $\pm$ 0.0009 & 0.128 $\pm$ 0.004 & 2.61 $\pm$ 0.01 & 0.98 & 991.3 / 1011 & 0.136 \\
16&P011466109503 & 2.17 $\pm$ 0.03 & 0.0337 $\pm$ 0.0018 & 0.091 $\pm$ 0.012 & 2.74 $\pm$ 0.03 & 0.92 & 678.3 / 741 & 0.140 \\
17&P011466109601 & 2.10 $\pm$ 0.02 & 0.0330 $\pm$ 0.0010 & 0.106 $\pm$ 0.005 & 2.67 $\pm$ 0.01 & 0.98 & 982.1 / 998 & 0.138 \\
18&P011466109602 & 2.09 $\pm$ 0.02 & 0.0344 $\pm$ 0.0012 & 0.106 $\pm$ 0.006 & 2.66 $\pm$ 0.02 & 1.05 & 961.7 / 918 & 0.137 \\
19&P011466109701 & 2.18 $\pm$ 0.02 & 0.0404 $\pm$ 0.0012 & 0.072 $\pm$ 0.008 & 2.70 $\pm$ 0.02 & 1.05 & 1002.3 / 957 & 0.136 \\
20&P011466109702 & 2.18 $\pm$ 0.02 & 0.0378 $\pm$ 0.0012 & 0.074 $\pm$ 0.008 & 2.69 $\pm$ 0.02 & 1.06 & 1011.1 / 950 & 0.136 \\
21&P011466109801 & 2.16 $\pm$ 0.03 & 0.0236 $\pm$ 0.0010 & 0.112 $\pm$ 0.005 & 2.53 $\pm$ 0.01 & 0.96 & 939.4 / 979 & 0.131 \\
22&P011466109802 & 2.21 $\pm$ 0.03 & 0.0268 $\pm$ 0.0011 & 0.107 $\pm$ 0.005 & 2.54 $\pm$ 0.01 & 1.04 & 1015.6 / 976 & 0.131 \\
23&P011466109803 & 2.24 $\pm$ 0.03 & 0.0247 $\pm$ 0.0012 & 0.115 $\pm$ 0.005 & 2.51 $\pm$ 0.01 & 1.01 & 966.9 / 960 & 0.130 \\
24&P011466109901 & 2.22 $\pm$ 0.03 & 0.0194 $\pm$ 0.0010 & 0.108 $\pm$ 0.004 & 2.50 $\pm$ 0.01 & 0.92 & 894.6 / 971 & 0.129 \\
25&P011466110001 & 2.50(f) & 0.0139 $\pm$ 0.0002 & 0.139 $\pm$ 0.004 & 2.36 $\pm$ 0.01 & 0.88 & 834.1 / 943 & 0.124 \\
26&P011466110101 & 2.50(f) & 0.0127 $\pm$ 0.0002 & 0.151 $\pm$ 0.004 & 2.28 $\pm$ 0.01 & 0.90 & 835.7 / 924 & 0.121 \\
27&P011466110201 & 2.50(f) & 0.0118 $\pm$ 0.0002 & 0.158 $\pm$ 0.004 & 2.21 $\pm$ 0.01 & 0.91 & 897.9 / 984 & 0.118 \\
28&P011466110301 & 2.50(f) & 0.0110 $\pm$ 0.0002 & 0.161 $\pm$ 0.005 & 2.18 $\pm$ 0.01 & 0.90 & 821.7 / 911 & 0.116 \\
29&P011466110302 & 2.50(f) & 0.0112 $\pm$ 0.0002 & 0.152 $\pm$ 0.004 & 2.20 $\pm$ 0.01 & 0.97 & 903.3 / 934 & 0.117 \\
30&P011466110401 & 2.50(f) & 0.0127 $\pm$ 0.0002 & 0.207 $\pm$ 0.003 & 2.04 $\pm$ 0.01 & 1.10 & 1040.2 / 947 & 0.112 \\
31&P011466110701 & 2.50(f) & 0.0113 $\pm$ 0.0002 & 0.187 $\pm$ 0.005 & 2.04 $\pm$ 0.01 & 1.04 & 1037.0 / 995 & 0.111 \\
32&P011466110801 & 2.50(f) & 0.0103 $\pm$ 0.0003 & 0.176 $\pm$ 0.006 & 2.05 $\pm$ 0.01 & 1.02 & 901.4 / 888 & 0.110 \\
33&P011466110901 & 2.50(f) & 0.0112 $\pm$ 0.0002 & 0.217 $\pm$ 0.003 & 1.89 $\pm$ 0.01 & 1.13 & 1064.2 / 945 & 0.104 \\
34&P011466110902 & 2.50(f) & 0.0121 $\pm$ 0.0004 & 0.218 $\pm$ 0.004 & 1.88 $\pm$ 0.01 & 0.90 & 716.3 / 800 & 0.104 \\
35&P011466111001 & 2.50(f) & 0.0146 $\pm$ 0.0003 & 0.223 $\pm$ 0.003 & 1.85 $\pm$ 0.01 & 1.13 & 1067.2 / 945 & 0.103 \\
36&P011466111201 & 2.50(f) & 0.0125 $\pm$ 0.0003 & 0.237 $\pm$ 0.004 & 1.75 $\pm$ 0.01 & 1.04 & 936.2 / 903 & 0.098 \\
37&P011466111301 & 2.50(f) & 0.0116 $\pm$ 0.0003 & 0.232 $\pm$ 0.004 & 1.74 $\pm$ 0.01 & 1.00 & 880.5 / 880 & 0.097 \\
38&P011466111401 & 2.50(f) & 0.0090 $\pm$ 0.0003 & 0.227 $\pm$ 0.003 & 1.69 $\pm$ 0.01 & 1.15 & 1034.1 / 900 & 0.094 \\
39&P011466111501 & 2.50(f) & 0.0105 $\pm$ 0.0003 & 0.242 $\pm$ 0.004 & 1.61 $\pm$ 0.01 & 1.26 & 1124.4 / 895 & 0.091 \\
40&P011466111601 & 2.50(f) & 0.0127 $\pm$ 0.0003 & 0.234 $\pm$ 0.004 & 1.57 $\pm$ 0.01 & 1.04 & 874.8 / 838 & 0.088 \\
41&P011466111701 & 2.50(f) & 0.0118 $\pm$ 0.0003 & 0.223 $\pm$ 0.004 & 1.54 $\pm$ 0.01 & 1.04 & 860.1 / 827 & 0.086 \\
42&P011466111801 & 2.50(f) & 0.0200 $\pm$ 0.0007 & 0.231 $\pm$ 0.007 & 1.47 $\pm$ 0.01 & 1.07 & 587.7 / 550 & 0.082 \\
43&P011466111802 & 2.50(f) & 0.0196 $\pm$ 0.0009 & 0.253 $\pm$ 0.014 & 1.41 $\pm$ 0.03 & 1.21 & 522.8 / 433 & 0.080 \\
44&P011466111901 & 2.50(f) & 0.0135 $\pm$ 0.0004 & 0.227 $\pm$ 0.006 & 1.41 $\pm$ 0.01 & 1.01 & 818.5 / 811 & 0.079 \\
45&P011466112001 & 2.50(f) & 0.0132 $\pm$ 0.0005 & 0.230 $\pm$ 0.005 & 1.37 $\pm$ 0.01 & 1.24 & 948.5 / 768 & 0.076 \\
46&P011466112101 & 2.50(f) & 0.0135 $\pm$ 0.0004 & 0.228 $\pm$ 0.005 & 1.32 $\pm$ 0.01 & 1.14 & 912.0 / 801 & 0.074 \\
47&P011466112201 & 2.50(f) & 0.0194 $\pm$ 0.0005 & 0.216 $\pm$ 0.006 & 1.30 $\pm$ 0.01 & 1.35 & 990.2 / 731 & 0.072 \\
48&P011466112301 & 2.50(f) & 0.0171 $\pm$ 0.0005 & 0.227 $\pm$ 0.009 & 1.24 $\pm$ 0.02 & 1.34 & 846.4 / 632 & 0.069 \\
49&P011466112401 & 2.50(f) & 0.0199 $\pm$ 0.0005 & 0.156 $\pm$ 0.009 & 1.15 $\pm$ 0.01 & 1.26 & 974.8 / 772 & 0.061 \\
50&P011466112402 & 2.50(f) & 0.0209 $\pm$ 0.0008 & 0.156 $\pm$ 0.015 & 1.14 $\pm$ 0.02 & 0.94 & 455.9 / 484 & 0.061 \\
51&P011466112501 & 2.50(f) & 0.0241 $\pm$ 0.0004 & 0.141 $\pm$ 0.007 & 1.14 $\pm$ 0.01 & 1.30 & 1095.5 / 843 & 0.060 \\
52&P011466112601 & 2.50(f) & 0.0227 $\pm$ 0.0004 & 0.149 $\pm$ 0.007 & 1.09 $\pm$ 0.01 & 1.16 & 933.6 / 804 & 0.058 \\
53&P011466112701 & 2.63 $\pm$ 0.04 & 0.0389 $\pm$ 0.0023 & 0.047 $\pm$ 0.014 & 1.15 $\pm$ 0.02 & 1.05 & 815.0 / 773 & 0.057 \\
54&P011466112801 & 2.32 $\pm$ 0.03 & 0.0491 $\pm$ 0.0019 & 0.016 $\pm$ 0.013 & 1.14 $\pm$ 0.02 & 0.98 & 771.3 / 791 & 0.056 \\
\hline
55&P011466108401 & 2.36 $\pm$ 0.01 & 0.2537 $\pm$ 0.0024 & 0.133 $\pm$ 0.009 & 2.83 $\pm$ 0.03 & 1.64 & 1850.6 / 1125 & 0.149 \\
56&P011466108402 & 2.34 $\pm$ 0.01 & 0.2521 $\pm$ 0.0024 & 0.153 $\pm$ 0.010 & 2.79 $\pm$ 0.03 & 1.68 & 1864.5 / 1108 & 0.148 \\
57&P011466108403 & 2.39 $\pm$ 0.01 & 0.2589 $\pm$ 0.0029 & 0.073 $\pm$ 0.014 & 3.02 $\pm$ 0.04 & 1.35 & 1412.7 / 1044 & 0.153 \\
58&P011466108501 & 2.31 $\pm$ 0.01 & 0.1479 $\pm$ 0.0022 & 0.121 $\pm$ 0.008 & 2.63 $\pm$ 0.02 & 1.50 & 1540.6 / 1024 & 0.137 \\
59&P011466112901 & 2.36 $\pm$ 0.01 & 0.1573 $\pm$ 0.0018 & 0.035 $\pm$ 0.015 & 1.07 $\pm$ 0.02 & 1.24 & 1292.8 / 1044 & 0.053 \\
60&P011466113001 & 2.29 $\pm$ 0.01 & 0.1875 $\pm$ 0.0020 & 0.109 $\pm$ 0.013 & 0.88 $\pm$ 0.01 & 1.18 & 1196.2 / 1013 & 0.046 \\
61&P011466113101 & 2.16 $\pm$ 0.01 & 0.2778 $\pm$ 0.0021 & 0.200 $\pm$ 0.022 & 0.60 $\pm$ 0.02 & 1.09 & 1183.6 / 1087 & 0.033 \\
\enddata
\tablecomments{$^a$ the dimensionless spin parameter a$_{*}$.\\
$^b$ the mass accretion rate $\dot{M}$ in units of 10$^{18}$ g s$^{-1}$.\\
$^c$ the bolometric Eddington-scaled luminosity $l = L(a_*,\dot M)/L_{\rm Edd}$.}
\end{deluxetable*}

\begin{deluxetable*}{lllrrrrrrll}
\tablecaption{Best-fit parameters for spectra with the model {\sc constant*tbabs*simpl*kerrbb2} \\($\alpha=0.1$, $M$ = $8.48~M_\odot$, $i=63^\circ$ and $D=2.96$ kpc)\label{alpha}}
\tablewidth{700pt}
\tabletypesize{\scriptsize}
\tablehead{
\colhead{Number} & \colhead{ObsID} & \multicolumn{2}{c}{\sc simpl} & \multicolumn{2}{c}{\sc kerrbb2}  & \colhead{Reduced $\chi^2_\nu$} & \colhead{$\chi^2$/d.o.f.}& \colhead{$l$} \\
\cline{3-6}
\colhead{} & \colhead{} &\colhead{$\Gamma$} & \colhead{$f_{\rm sc}$} & \colhead{$a_{*}$$^a$} & \colhead{$\dot{M}$$^b$} 
} 
\startdata
1&P011466108502 & 2.25 $\pm$ 0.01 & 0.1030 $\pm$ 0.0021 & 0.059 $\pm$ 0.011 & 2.86 $\pm$ 0.03 & 1.22 & 1129.4 / 923 & 0.143 \\
2&P011466108601 & 2.27 $\pm$ 0.02 & 0.0882 $\pm$ 0.0022 & 0.029 $\pm$ 0.010 & 3.04 $\pm$ 0.03 & 1.18 & 1087.8 / 921 & 0.150 \\
3&P011466108602 & 2.15 $\pm$ 0.01 & 0.0963 $\pm$ 0.0019 & 0.046 $\pm$ 0.011 & 2.94 $\pm$ 0.03 & 1.10 & 974.6 / 887 & 0.146 \\
4&P011466108702 & 2.16 $\pm$ 0.02 & 0.0563 $\pm$ 0.0018 & 0.036 $\pm$ 0.010 & 3.01 $\pm$ 0.03 & 1.09 & 914.4 / 835 & 0.149 \\
5&P011466108801 & 2.14 $\pm$ 0.02 & 0.0465 $\pm$ 0.0013 & 0.019 $\pm$ 0.006 & 3.04 $\pm$ 0.02 & 1.06 & 1037.5 / 980 & 0.149 \\
6&P011466108802 & 2.07 $\pm$ 0.02 & 0.0405 $\pm$ 0.0014 & 0.036 $\pm$ 0.008 & 2.99 $\pm$ 0.02 & 1.01 & 910.4 / 899 & 0.148 \\
7&P011466108901 & 2.05 $\pm$ 0.02 & 0.0265 $\pm$ 0.0010 & 0.098 $\pm$ 0.009 & 2.73 $\pm$ 0.02 & 0.95 & 836.0 / 878 & 0.140 \\
8&P011466109001 & 2.08 $\pm$ 0.02 & 0.0309 $\pm$ 0.0008 & 0.055 $\pm$ 0.006 & 2.88 $\pm$ 0.02 & 1.11 & 1105.2 / 993 & 0.144 \\
9&P011466109101 & 2.17 $\pm$ 0.02 & 0.0430 $\pm$ 0.0016 & 0.015 $\pm$ 0.007 & 2.99 $\pm$ 0.02 & 0.95 & 856.6 / 903 & 0.146 \\
10&P011466109102 & 2.07 $\pm$ 0.02 & 0.0348 $\pm$ 0.0011 & 0.039 $\pm$ 0.007 & 2.91 $\pm$ 0.02 & 0.98 & 904.8 / 923 & 0.144 \\
11&P011466109201 & 2.08 $\pm$ 0.02 & 0.0330 $\pm$ 0.0011 & 0.058 $\pm$ 0.007 & 2.82 $\pm$ 0.02 & 0.99 & 891.9 / 901 & 0.141 \\
12&P011466109202 & 2.09 $\pm$ 0.03 & 0.0311 $\pm$ 0.0012 & 0.062 $\pm$ 0.007 & 2.80 $\pm$ 0.02 & 1.00 & 922.6 / 921 & 0.141 \\
13&P011466109301 & 2.05 $\pm$ 0.02 & 0.0251 $\pm$ 0.0009 & 0.061 $\pm$ 0.007 & 2.79 $\pm$ 0.02 & 1.03 & 913.6 / 888 & 0.140 \\
14&P011466109401 & 2.07 $\pm$ 0.03 & 0.0212 $\pm$ 0.0008 & 0.077 $\pm$ 0.007 & 2.70 $\pm$ 0.02 & 1.02 & 911.8 / 896 & 0.137 \\
15&P011466109501 & 2.15 $\pm$ 0.02 & 0.0242 $\pm$ 0.0008 & 0.065 $\pm$ 0.006 & 2.72 $\pm$ 0.01 & 0.98 & 992.3 / 1011 & 0.137 \\
16&P011466109503 & 2.15 $\pm$ 0.04 & 0.0331 $\pm$ 0.0019 & 0.031 $\pm$ 0.010 & 2.84 $\pm$ 0.03 & 0.91 & 677.4 / 741 & 0.140 \\
17&P011466109601 & 2.10 $\pm$ 0.02 & 0.0333 $\pm$ 0.0009 & 0.037 $\pm$ 0.006 & 2.79 $\pm$ 0.02 & 0.98 & 976.5 / 998 & 0.138 \\
18&P011466109602 & 2.10 $\pm$ 0.02 & 0.0348 $\pm$ 0.0012 & 0.037 $\pm$ 0.007 & 2.78 $\pm$ 0.02 & 1.04 & 957.4 / 918 & 0.138 \\
19&P011466109701 & 2.17 $\pm$ 0.02 & 0.0397 $\pm$ 0.0012 & 0.018 $\pm$ 0.006 & 2.79 $\pm$ 0.02 & 1.05 & 1001.2 / 957 & 0.136 \\
20&P011466109702 & 2.16 $\pm$ 0.02 & 0.0372 $\pm$ 0.0012 & 0.020 $\pm$ 0.006 & 2.77 $\pm$ 0.02 & 1.06 & 1010.0 / 950 & 0.136 \\
21&P011466109801 & 2.17 $\pm$ 0.03 & 0.0240 $\pm$ 0.0010 & 0.046 $\pm$ 0.006 & 2.64 $\pm$ 0.01 & 0.96 & 935.9 / 979 & 0.132 \\
22&P011466109802 & 2.22 $\pm$ 0.02 & 0.0271 $\pm$ 0.0010 & 0.040 $\pm$ 0.006 & 2.65 $\pm$ 0.01 & 1.03 & 1009.8 / 976 & 0.131 \\
23&P011466109803 & 2.25 $\pm$ 0.03 & 0.0251 $\pm$ 0.0011 & 0.050 $\pm$ 0.006 & 2.63 $\pm$ 0.02 & 1.00 & 964.1 / 960 & 0.131 \\
24&P011466109901 & 2.23 $\pm$ 0.03 & 0.0196 $\pm$ 0.0009 & 0.042 $\pm$ 0.005 & 2.62 $\pm$ 0.01 & 0.92 & 890.3 / 971 & 0.130 \\
25&P011466110001 & 2.50(f) & 0.0141 $\pm$ 0.0002 & 0.083 $\pm$ 0.005 & 2.45 $\pm$ 0.01 & 0.89 & 838.4 / 943 & 0.125 \\
26&P011466110101 & 2.50(f) & 0.0129 $\pm$ 0.0002 & 0.102 $\pm$ 0.003 & 2.35 $\pm$ 0.01 & 0.91 & 839.5 / 924 & 0.121 \\
27&P011466110201 & 2.50(f) & 0.0119 $\pm$ 0.0002 & 0.109 $\pm$ 0.003 & 2.28 $\pm$ 0.01 & 0.92 & 901.7 / 984 & 0.118 \\
28&P011466110301 & 2.50(f) & 0.0111 $\pm$ 0.0002 & 0.111 $\pm$ 0.003 & 2.25 $\pm$ 0.01 & 0.90 & 823.5 / 911 & 0.116 \\
29&P011466110302 & 2.50(f) & 0.0113 $\pm$ 0.0003 & 0.105 $\pm$ 0.003 & 2.27 $\pm$ 0.01 & 0.97 & 909.2 / 934 & 0.117 \\
30&P011466110401 & 2.50(f) & 0.0129 $\pm$ 0.0002 & 0.142 $\pm$ 0.004 & 2.14 $\pm$ 0.01 & 1.08 & 1026.1 / 947 & 0.113 \\
31&P011466110701 & 2.50(f) & 0.0113 $\pm$ 0.0002 & 0.132 $\pm$ 0.003 & 2.12 $\pm$ 0.01 & 1.02 & 1018.1 / 995 & 0.111 \\
32&P011466110801 & 2.50(f) & 0.0103 $\pm$ 0.0003 & 0.124 $\pm$ 0.004 & 2.12 $\pm$ 0.01 & 1.01 & 897.7 / 888 & 0.110 \\
33&P011466110901 & 2.50(f) & 0.0115 $\pm$ 0.0002 & 0.156 $\pm$ 0.004 & 1.98 $\pm$ 0.01 & 1.11 & 1053.4 / 945 & 0.105 \\
34&P011466110902 & 2.50(f) & 0.0124 $\pm$ 0.0004 & 0.160 $\pm$ 0.006 & 1.97 $\pm$ 0.01 & 0.89 & 714.5 / 800 & 0.105 \\
35&P011466111001 & 2.50(f) & 0.0150 $\pm$ 0.0003 & 0.164 $\pm$ 0.004 & 1.94 $\pm$ 0.01 & 1.12 & 1057.4 / 945 & 0.104 \\
36&P011466111201 & 2.50(f) & 0.0129 $\pm$ 0.0003 & 0.186 $\pm$ 0.005 & 1.82 $\pm$ 0.01 & 1.04 & 941.3 / 903 & 0.099 \\
37&P011466111301 & 2.50(f) & 0.0120 $\pm$ 0.0003 & 0.177 $\pm$ 0.005 & 1.81 $\pm$ 0.01 & 1.00 & 882.2 / 880 & 0.098 \\
38&P011466111401 & 2.50(f) & 0.0093 $\pm$ 0.0003 & 0.171 $\pm$ 0.005 & 1.76 $\pm$ 0.01 & 1.15 & 1030.6 / 900 & 0.095 \\
39&P011466111501 & 2.50(f) & 0.0107 $\pm$ 0.0003 & 0.206 $\pm$ 0.003 & 1.65 $\pm$ 0.01 & 1.26 & 1131.8 / 895 & 0.091 \\
40&P011466111601 & 2.50(f) & 0.0131 $\pm$ 0.0003 & 0.180 $\pm$ 0.007 & 1.63 $\pm$ 0.01 & 1.04 & 874.2 / 838 & 0.088 \\
41&P011466111701 & 2.50(f) & 0.0121 $\pm$ 0.0003 & 0.167 $\pm$ 0.006 & 1.60 $\pm$ 0.01 & 1.04 & 856.9 / 827 & 0.086 \\
42&P011466111801 & 2.50(f) & 0.0206 $\pm$ 0.0007 & 0.177 $\pm$ 0.011 & 1.53 $\pm$ 0.02 & 1.06 & 584.7 / 550 & 0.083 \\
43&P011466111802 & 2.50(f) & 0.0200 $\pm$ 0.0009 & 0.216 $\pm$ 0.012 & 1.44 $\pm$ 0.02 & 1.20 & 520.3 / 433 & 0.080 \\
44&P011466111901 & 2.50(f) & 0.0140 $\pm$ 0.0004 & 0.170 $\pm$ 0.009 & 1.48 $\pm$ 0.01 & 1.00 & 810.4 / 811 & 0.079 \\
45&P011466112001 & 2.50(f) & 0.0138 $\pm$ 0.0004 & 0.176 $\pm$ 0.008 & 1.43 $\pm$ 0.01 & 1.23 & 942.1 / 768 & 0.077 \\
46&P011466112101 & 2.50(f) & 0.0141 $\pm$ 0.0004 & 0.172 $\pm$ 0.008 & 1.38 $\pm$ 0.01 & 1.13 & 904.0 / 801 & 0.074 \\
47&P011466112201 & 2.50(f) & 0.0215 $\pm$ 0.0005 & 0.145 $\pm$ 0.008 & 1.38 $\pm$ 0.01 & 1.44 & 1055.0 / 734 & 0.073 \\
48&P011466112301 & 2.50(f) & 0.0178 $\pm$ 0.0006 & 0.166 $\pm$ 0.014 & 1.31 $\pm$ 0.02 & 1.33 & 837.6 / 632 & 0.070 \\
49&P011466112401 & 2.50(f) & 0.0203 $\pm$ 0.0005 & 0.114 $\pm$ 0.007 & 1.18 $\pm$ 0.01 & 1.24 & 954.9 / 772 & 0.061 \\
50&P011466112402 & 2.50(f) & 0.0214 $\pm$ 0.0008 & 0.111 $\pm$ 0.011 & 1.17 $\pm$ 0.02 & 0.93 & 450.5 / 484 & 0.061 \\
51&P011466112501 & 2.50(f) & 0.0248 $\pm$ 0.0004 & 0.083 $\pm$ 0.009 & 1.18 $\pm$ 0.01 & 1.26 & 1062.1 / 843 & 0.060 \\
52&P011466112601 & 2.50(f) & 0.0234 $\pm$ 0.0004 & 0.092 $\pm$ 0.010 & 1.13 $\pm$ 0.01 & 1.14 & 914.8 / 804 & 0.058 \\
53&P011466112701 & 2.59 $\pm$ 0.04 & 0.0377 $\pm$ 0.0020 & -0.001 $\pm$ 0.015 & 1.18 $\pm$ 0.02 & 1.05 & 809.4 / 773 & 0.057 \\
54&P011466112801 & 2.30 $\pm$ 0.02 & 0.0480 $\pm$ 0.0016 & -0.038 $\pm$ 0.014 & 1.18 $\pm$ 0.02 & 0.97 & 766.4 / 791 & 0.056 \\
\hline
55&P011466108401 & 2.36 $\pm$ 0.01 & 0.2556 $\pm$ 0.0023 & 0.058 $\pm$ 0.011 & 2.99 $\pm$ 0.03 & 1.64 & 1847.8 / 1125 & 0.150 \\
56&P011466108402 & 2.35 $\pm$ 0.01 & 0.2547 $\pm$ 0.0023 & 0.079 $\pm$ 0.012 & 2.94 $\pm$ 0.03 & 1.68 & 1862.2 / 1108 & 0.149 \\
57&P011466108403 & 2.38 $\pm$ 0.01 & 0.2551 $\pm$ 0.0031 & 0.028 $\pm$ 0.013 & 3.09 $\pm$ 0.03 & 1.35 & 1404.4 / 1044 & 0.152 \\
58&P011466108501 & 2.32 $\pm$ 0.01 & 0.1497 $\pm$ 0.0021 & 0.047 $\pm$ 0.010 & 2.77 $\pm$ 0.02 & 1.49 & 1530.4 / 1024 & 0.138 \\
59&P011466112901 & 2.34 $\pm$ 0.01 & 0.1541 $\pm$ 0.0016 & -0.010 $\pm$ 0.016 & 1.09 $\pm$ 0.02 & 1.21 & 1258.2 / 1044 & 0.053 \\
60&P011466113001 & 2.27 $\pm$ 0.01 & 0.1841 $\pm$ 0.0018 & 0.041 $\pm$ 0.016 & 0.92 $\pm$ 0.02 & 1.15 & 1165.5 / 1013 & 0.046 \\
61&P011466113101 & 2.16 $\pm$ 0.00 & 0.2747 $\pm$ 0.0022 & 0.058 $\pm$ 0.024 & 0.68 $\pm$ 0.02 & 1.08 & 1172.2 / 1087 & 0.034 \\
\enddata
\end{deluxetable*}

\begin{deluxetable*}{lllrrrrrrll}
\tablecaption{Best-fit parameters for spectra with the model {\sc constant*tbabs*simpl*kerrbb2} \\($\alpha=0.01$, $M=8.06~M_\odot$, $i=66\fdg2$ and $D=2.96$ kpc)\label{torres}}
\tablewidth{700pt}
\tabletypesize{\scriptsize}
\tablehead{
\colhead{Number} & \colhead{ObsID} & \multicolumn{2}{c}{\sc simpl} & \multicolumn{2}{c}{\sc kerrbb2}   & \colhead{Reduced $\chi^2_\nu$} & \colhead{$\chi^2$/d.o.f.}& \colhead{$l$}\\
\cline{3-6}
\colhead{} & \colhead{} &\colhead{$\Gamma$} & \colhead{$f_{\rm sc}$} & \colhead{$a_{*}$$^a$} & \colhead{$\dot{M}$$^b$} 
} 
\startdata
1&P011466108502 & 2.25 $\pm$ 0.01 & 0.1046 $\pm$ 0.0021 & -0.129 $\pm$ 0.013 & 3.63 $\pm$ 0.04 & 1.23 & 1137.9 / 923 & 0.173 \\
2&P011466108601 & 2.28 $\pm$ 0.02 & 0.0897 $\pm$ 0.0022 & -0.163 $\pm$ 0.011 & 3.86 $\pm$ 0.04 & 1.19 & 1093.1 / 921 & 0.181 \\
3&P011466108602 & 2.16 $\pm$ 0.01 & 0.0974 $\pm$ 0.0019 & -0.141 $\pm$ 0.012 & 3.73 $\pm$ 0.04 & 1.10 & 979.3 / 887 & 0.177 \\
4&P011466108702 & 2.16 $\pm$ 0.02 & 0.0572 $\pm$ 0.0018 & -0.152 $\pm$ 0.011 & 3.82 $\pm$ 0.04 & 1.10 & 918.3 / 835 & 0.180 \\
5&P011466108801 & 2.15 $\pm$ 0.02 & 0.0474 $\pm$ 0.0013 & -0.172 $\pm$ 0.007 & 3.86 $\pm$ 0.02 & 1.06 & 1043.0 / 980 & 0.180 \\
6&P011466108802 & 2.08 $\pm$ 0.02 & 0.0415 $\pm$ 0.0015 & -0.153 $\pm$ 0.009 & 3.79 $\pm$ 0.03 & 1.02 & 916.9 / 899 & 0.178 \\
7&P011466108901 & 2.04 $\pm$ 0.02 & 0.0262 $\pm$ 0.0010 & -0.079 $\pm$ 0.007 & 3.45 $\pm$ 0.02 & 0.95 & 836.3 / 878 & 0.168 \\
8&P011466109001 & 2.09 $\pm$ 0.02 & 0.0318 $\pm$ 0.0009 & -0.129 $\pm$ 0.007 & 3.65 $\pm$ 0.02 & 1.12 & 1116.2 / 993 & 0.174 \\
9&P011466109101 & 2.17 $\pm$ 0.02 & 0.0439 $\pm$ 0.0016 & -0.177 $\pm$ 0.009 & 3.80 $\pm$ 0.03 & 0.95 & 861.1 / 903 & 0.177 \\
10&P011466109102 & 2.08 $\pm$ 0.02 & 0.0355 $\pm$ 0.0011 & -0.148 $\pm$ 0.008 & 3.69 $\pm$ 0.02 & 0.99 & 912.2 / 923 & 0.174 \\
11&P011466109201 & 2.09 $\pm$ 0.02 & 0.0339 $\pm$ 0.0011 & -0.126 $\pm$ 0.009 & 3.57 $\pm$ 0.03 & 1.00 & 898.7 / 901 & 0.170 \\
12&P011466109202 & 2.10 $\pm$ 0.02 & 0.0320 $\pm$ 0.0012 & -0.122 $\pm$ 0.009 & 3.55 $\pm$ 0.03 & 1.01 & 929.1 / 921 & 0.170 \\
13&P011466109301 & 2.06 $\pm$ 0.02 & 0.0257 $\pm$ 0.0009 & -0.122 $\pm$ 0.008 & 3.53 $\pm$ 0.02 & 1.04 & 921.4 / 888 & 0.169 \\
14&P011466109401 & 2.06 $\pm$ 0.03 & 0.0211 $\pm$ 0.0009 & -0.095 $\pm$ 0.006 & 3.39 $\pm$ 0.02 & 1.02 & 918.1 / 896 & 0.164 \\
15&P011466109501 & 2.17 $\pm$ 0.02 & 0.0250 $\pm$ 0.0008 & -0.118 $\pm$ 0.007 & 3.45 $\pm$ 0.02 & 0.99 & 1002.1 / 1011 & 0.165 \\
16&P011466109503 & 2.16 $\pm$ 0.04 & 0.0338 $\pm$ 0.0019 & -0.159 $\pm$ 0.012 & 3.60 $\pm$ 0.04 & 0.92 & 679.1 / 741 & 0.169 \\
17&P011466109601 & 2.11 $\pm$ 0.02 & 0.0341 $\pm$ 0.0009 & -0.152 $\pm$ 0.007 & 3.54 $\pm$ 0.02 & 0.99 & 985.2 / 998 & 0.167 \\
18&P011466109602 & 2.11 $\pm$ 0.02 & 0.0356 $\pm$ 0.0012 & -0.153 $\pm$ 0.008 & 3.53 $\pm$ 0.02 & 1.05 & 963.9 / 918 & 0.166 \\
19&P011466109701 & 2.18 $\pm$ 0.02 & 0.0405 $\pm$ 0.0012 & -0.176 $\pm$ 0.007 & 3.54 $\pm$ 0.02 & 1.05 & 1006.0 / 957 & 0.165 \\
20&P011466109702 & 2.17 $\pm$ 0.02 & 0.0379 $\pm$ 0.0012 & -0.174 $\pm$ 0.007 & 3.52 $\pm$ 0.02 & 1.07 & 1015.4 / 950 & 0.164 \\
21&P011466109801 & 2.18 $\pm$ 0.03 & 0.0247 $\pm$ 0.0010 & -0.142 $\pm$ 0.007 & 3.35 $\pm$ 0.02 & 0.96 & 942.4 / 979 & 0.159 \\
22&P011466109802 & 2.23 $\pm$ 0.02 & 0.0278 $\pm$ 0.0010 & -0.150 $\pm$ 0.007 & 3.36 $\pm$ 0.02 & 1.04 & 1016.6 / 976 & 0.158 \\
23&P011466109803 & 2.26 $\pm$ 0.03 & 0.0260 $\pm$ 0.0012 & -0.138 $\pm$ 0.008 & 3.33 $\pm$ 0.02 & 1.01 & 970.1 / 960 & 0.158 \\
24&P011466109901 & 2.24 $\pm$ 0.03 & 0.0203 $\pm$ 0.0009 & -0.146 $\pm$ 0.007 & 3.32 $\pm$ 0.02 & 0.92 & 893.8 / 971 & 0.156 \\
25&P011466110001 & 2.50(f) & 0.0142 $\pm$ 0.0002 & -0.093 $\pm$ 0.004 & 3.09 $\pm$ 0.01 & 0.90 & 844.1 / 943 & 0.150 \\
26&P011466110101 & 2.50(f) & 0.0129 $\pm$ 0.0002 & -0.082 $\pm$ 0.004 & 2.99 $\pm$ 0.01 & 0.91 & 838.6 / 924 & 0.146 \\
27&P011466110201 & 2.50(f) & 0.0120 $\pm$ 0.0002 & -0.073 $\pm$ 0.004 & 2.90 $\pm$ 0.01 & 0.92 & 904.2 / 984 & 0.142 \\
28&P011466110301 & 2.50(f) & 0.0112 $\pm$ 0.0002 & -0.071 $\pm$ 0.004 & 2.86 $\pm$ 0.01 & 0.90 & 824.4 / 911 & 0.140 \\
29&P011466110302 & 2.50(f) & 0.0114 $\pm$ 0.0003 & -0.079 $\pm$ 0.004 & 2.88 $\pm$ 0.01 & 0.97 & 909.7 / 934 & 0.141 \\
30&P011466110401 & 2.50(f) & 0.0130 $\pm$ 0.0002 & -0.031 $\pm$ 0.005 & 2.71 $\pm$ 0.01 & 1.10 & 1040.3 / 947 & 0.136 \\
31&P011466110701 & 2.50(f) & 0.0114 $\pm$ 0.0002 & -0.044 $\pm$ 0.004 & 2.69 $\pm$ 0.01 & 1.04 & 1032.8 / 995 & 0.134 \\
32&P011466110801 & 2.50(f) & 0.0104 $\pm$ 0.0003 & -0.055 $\pm$ 0.005 & 2.69 $\pm$ 0.01 & 1.02 & 902.7 / 888 & 0.133 \\
33&P011466110901 & 2.50(f) & 0.0116 $\pm$ 0.0002 & -0.013 $\pm$ 0.006 & 2.50 $\pm$ 0.01 & 1.14 & 1078.3 / 945 & 0.127 \\
34&P011466110902 & 2.50(f) & 0.0125 $\pm$ 0.0004 & -0.007 $\pm$ 0.008 & 2.48 $\pm$ 0.02 & 0.90 & 721.1 / 800 & 0.126 \\
35&P011466111001 & 2.50(f) & 0.0149 $\pm$ 0.0003 & 0.008 $\pm$ 0.003 & 2.42 $\pm$ 0.01 & 1.15 & 1082.7 / 945 & 0.124 \\
36&P011466111201 & 2.50(f) & 0.0127 $\pm$ 0.0003 & 0.024 $\pm$ 0.004 & 2.28 $\pm$ 0.01 & 1.05 & 944.2 / 903 & 0.118 \\
37&P011466111301 & 2.50(f) & 0.0118 $\pm$ 0.0003 & 0.017 $\pm$ 0.004 & 2.27 $\pm$ 0.01 & 1.01 & 890.4 / 880 & 0.117 \\
38&P011466111401 & 2.50(f) & 0.0092 $\pm$ 0.0003 & 0.011 $\pm$ 0.004 & 2.20 $\pm$ 0.01 & 1.16 & 1045.7 / 900 & 0.113 \\
39&P011466111501 & 2.50(f) & 0.0108 $\pm$ 0.0003 & 0.028 $\pm$ 0.004 & 2.11 $\pm$ 0.01 & 1.26 & 1131.7 / 895 & 0.109 \\
40&P011466111601 & 2.50(f) & 0.0130 $\pm$ 0.0003 & 0.016 $\pm$ 0.005 & 2.05 $\pm$ 0.01 & 1.05 & 880.8 / 838 & 0.106 \\
41&P011466111701 & 2.50(f) & 0.0123 $\pm$ 0.0003 & -0.008 $\pm$ 0.008 & 2.04 $\pm$ 0.02 & 1.04 & 863.2 / 827 & 0.103 \\
42&P011466111801 & 2.50(f) & 0.0205 $\pm$ 0.0007 & 0.010 $\pm$ 0.009 & 1.93 $\pm$ 0.02 & 1.07 & 587.4 / 550 & 0.099 \\
43&P011466111802 & 2.50(f) & 0.0201 $\pm$ 0.0009 & 0.036 $\pm$ 0.017 & 1.85 $\pm$ 0.03 & 1.20 & 521.4 / 433 & 0.096 \\
44&P011466111901 & 2.50(f) & 0.0139 $\pm$ 0.0004 & 0.008 $\pm$ 0.007 & 1.85 $\pm$ 0.01 & 1.01 & 818.5 / 811 & 0.095 \\
45&P011466112001 & 2.50(f) & 0.0136 $\pm$ 0.0005 & 0.011 $\pm$ 0.006 & 1.79 $\pm$ 0.01 & 1.24 & 948.8 / 768 & 0.092 \\
46&P011466112101 & 2.50(f) & 0.0139 $\pm$ 0.0004 & 0.009 $\pm$ 0.006 & 1.73 $\pm$ 0.01 & 1.14 & 914.8 / 801 & 0.089 \\
47&P011466112201 & 2.50(f) & 0.0215 $\pm$ 0.0005 & -0.042 $\pm$ 0.011 & 1.76 $\pm$ 0.02 & 1.45 & 1064.9 / 734 & 0.088 \\
48&P011466112301 & 2.50(f) & 0.0178 $\pm$ 0.0006 & -0.016 $\pm$ 0.019 & 1.67 $\pm$ 0.03 & 1.34 & 844.2 / 632 & 0.084 \\
49&P011466112401 & 2.50(f) & 0.0202 $\pm$ 0.0005 & -0.080 $\pm$ 0.008 & 1.51 $\pm$ 0.01 & 1.25 & 966.9 / 772 & 0.074 \\
50&P011466112402 & 2.50(f) & 0.0213 $\pm$ 0.0008 & -0.083 $\pm$ 0.014 & 1.50 $\pm$ 0.02 & 0.94 & 453.8 / 484 & 0.073 \\
51&P011466112501 & 2.50(f) & 0.0247 $\pm$ 0.0004 & -0.113 $\pm$ 0.011 & 1.51 $\pm$ 0.01 & 1.29 & 1083.6 / 843 & 0.072 \\
52&P011466112601 & 2.50(f) & 0.0230 $\pm$ 0.0004 & -0.092 $\pm$ 0.007 & 1.42 $\pm$ 0.01 & 1.16 & 929.8 / 804 & 0.069 \\
53&P011466112701 & 2.63 $\pm$ 0.04 & 0.0392 $\pm$ 0.0022 & -0.232 $\pm$ 0.018 & 1.53 $\pm$ 0.02 & 1.05 & 812.1 / 773 & 0.069 \\
54&P011466112801 & 2.32 $\pm$ 0.03 & 0.0495 $\pm$ 0.0018 & -0.276 $\pm$ 0.016 & 1.52 $\pm$ 0.02 & 0.97 & 769.5 / 791 & 0.067 \\
\hline
55&P011466108401 & 2.36 $\pm$ 0.01 & 0.2581 $\pm$ 0.0022 & -0.134 $\pm$ 0.012 & 3.81 $\pm$ 0.04 & 1.66 & 1864.2 / 1125 & 0.181 \\
56&P011466108402 & 2.34 $\pm$ 0.01 & 0.2523 $\pm$ 0.0024 & -0.080 $\pm$ 0.011 & 3.67 $\pm$ 0.03 & 1.69 & 1877.8 / 1108 & 0.179 \\
57&P011466108403 & 2.38 $\pm$ 0.01 & 0.2578 $\pm$ 0.0031 & -0.170 $\pm$ 0.015 & 3.96 $\pm$ 0.04 & 1.35 & 1411.2 / 1044 & 0.185 \\
58&P011466108501 & 2.32 $\pm$ 0.01 & 0.1518 $\pm$ 0.0021 & -0.146 $\pm$ 0.012 & 3.53 $\pm$ 0.03 & 1.51 & 1542.9 / 1024 & 0.167 \\
59&P011466112901 & 2.36 $\pm$ 0.01 & 0.1588 $\pm$ 0.0018 & -0.265 $\pm$ 0.018 & 1.43 $\pm$ 0.02 & 1.24 & 1290.7 / 1044 & 0.064 \\
60&P011466113001 & 2.29 $\pm$ 0.01 & 0.1884 $\pm$ 0.0020 & -0.167 $\pm$ 0.022 & 1.18 $\pm$ 0.02 & 1.18 & 1194.8 / 1013 & 0.055 \\
61&P011466113101 & 2.17 $\pm$ 0.01 & 0.2802 $\pm$ 0.0022 & -0.146 $\pm$ 0.035 & 0.87 $\pm$ 0.03 & 1.09 & 1189.3 / 1087 & 0.041 \\
\enddata
\end{deluxetable*}

\begin{deluxetable*}{lllrrrrrrll}
\tablecaption{Effect of Different $\Gamma$ and {\sc constant} normalization \\(only for P011466110701, where $\alpha=0.01$, $M$ = $8.48~M_\odot$, $i=63^\circ$ and $D=2.96$ kpc)\label{gamma}}
\tablewidth{700pt}
\tabletypesize{\scriptsize}
\tablehead{
\colhead{N} & \colhead{Model} & \colhead{Parameter} & \colhead{Case 1} & \colhead{Case 2} & \colhead{Case 3} & \colhead{Case 4} & \colhead{Case 5} & \colhead{Case 6}& \colhead{Case 7}
}

\startdata
1	&	{\sc simpl}	&	$\Gamma$	&	2.10(f)			&	2.30(f)			&	2.50(f)			&	2.50(f)			&	2.50(f)			&	2.70(f)			&	2.90(f)			\\
2	&	{\sc simpl}	&	$f_{\rm sc}$	&	0.0058	$\pm$	0.0001	&	0.0081	$\pm$	0.0002	&	0.0113	$\pm$	0.0002	&	0.0108	$\pm$	0.0002	&	0.0101	$\pm$	0.0002	&	0.0152	$\pm$	0.0003	&	0.0201	$\pm$	0.0004	\\
3	&	{\sc kerrbb2}	&	$a_*$	&	0.213	$\pm$	0.002	&	0.207	$\pm$	0.002	&	0.187	$\pm$	0.005	&	0.190	$\pm$	0.005	&	0.204	$\pm$	0.002	&	0.174	$\pm$	0.004	&	0.160	$\pm$	0.004	\\
4	&	{\sc kerrbb2}	&	$\dot{M}$	&	1.99	$\pm$	0.01	&	2.00	$\pm$	0.01	&	2.04	$\pm$	0.01	&	2.04	$\pm$	0.01	&	2.01	$\pm$	0.01	&	2.07	$\pm$	0.01	&	2.10	$\pm$	0.01	\\
5	&	{\sc constant}	&	norm	&	0.95(f)			&	0.95(f)			&	0.95(f)			&	1.00(f)			&	1.05(f)			&	0.95(f)			&	0.95(f)			\\
6	&		&	Reduced $\chi^2_\nu$	&	1.21			&	1.12			&	1.04			&	1.07			&	1.11			&	0.97			&	0.91			\\
7	&		&	$\chi^2$/d.o.f.	&	1201.1	/	995	&	1118.3	/	995	&	1037.0	/	995	&	1069.5	/	995	&	1100.6	/	995	&	963.5	/	995	&	908.6	/	995	\\
8	&		&	{$l$}	&	0.110			&	0.110			&	0.111			&	0.111			&	0.110			&	0.112			&	0.112			\\
\enddata
\end{deluxetable*}

\acknowledgments
 This work made use of the data from the \emph{Insight}-HXMT mission, a project funded by China National Space Administration (CNSA) and the Chinese Academy of Sciences (CAS). The authors thank supports from the National Program on Key Research and Development Project (Grant No. 2016YFA0400800 and 2016YFA0400801) and from the NSFC (U1838201 and U1838202). L.J.G. acknowledges the support by the National Program on Key Research and Development Project (Grant No. 2016YFA0400804), and by the NSFC (U1838114), and by the Strategic Priority Research Program of the Chinese Academy of Sciences (XDB23040100).

\newpage
\bibliography{ref}{}
\bibliographystyle{aasjournal}

\end{document}